\newif\ifcol
\coltrue

\newif\ifbw
\bwfalse

\newif\ifdraft
\draftfalse

\newcommand{\pagesize}{a4paper}

\ifdraft
  \colfalse
\fi

\ifdraft
  \documentclass[11pt,draftcls,\pagesize,onecolumn,twoside]{IEEEtran}
\else
  \documentclass[10pt,journal,\pagesize,twoside]{IEEEtran}
\fi

\usepackage{cite}      

\usepackage{graphicx}  
                        
\usepackage{subfigure} 
                        
\usepackage{url}       
\usepackage{ifpdf}
\ifpdf
  \usepackage[pdftex]{hyperref}
\else
  \usepackage[ps2pdf]{hyperref}
\fi                        
\usepackage{amsmath}
\usepackage{amsfonts}
\usepackage{amssymb}
\usepackage{color}

\newcommand{\vect}[1]{\ensuremath{\mbox{\boldmath ${#1}$}}}

\newcommand{\bl}[1]{\textcolor{black}{#1}}

\title{Exact Wavelets on the Ball}

\begin{document}

\author{Boris Leistedt and Jason D. McEwen
	\thanks{We gratefully acknowledge use of software to access seismological Earth data made available on Frederik Simons' webpage: {http://www.frederik.net}.  BL is supported by the Perren Fund and the Impact Fund (e-mail: boris.leistedt@ucl.ac.uk). JDM is supported by a Newton International Fellowship from the Royal Society and the British Academy and, during this work, was also supported by a Leverhulme Early Career Fellowship from the Leverhulme Trust (e-mail: jason.mcewen@ucl.ac.uk).}%
	\thanks{The authors are with the Department of Physics and Astronomy, University College London, London WC1E 6BT, United-Kingdom.}
}

\markboth{IEEE Transactions on Signal Processing,~Vol.~60, No.~12,~December~2012}%
{Leistedt \& McEwen: Exact wavelets on the ball}

\maketitle

\begin{abstract}  
We develop an exact wavelet transform on the three-dimensional ball (\emph{i.e.} on the solid sphere), which we name the \emph{flaglet} transform. For this purpose we first construct an exact transform on the \bl{radial half-line} using damped Laguerre polynomials and develop a corresponding quadrature rule. Combined with the spherical harmonic transform, this approach leads to a sampling theorem on the ball and a novel three-dimensional decomposition which we call the Fourier-Laguerre transform. We relate this new transform to the well-known Fourier-Bessel decomposition and show that \bl{band-limitedness} in the Fourier-Laguerre basis is a sufficient condition to compute the Fourier-Bessel decomposition exactly. We then construct the flaglet transform on the ball through a harmonic tiling, which is exact thanks to the exactness of the Fourier-Laguerre transform (from which the name flaglets is coined). The corresponding wavelet kernels are \bl{well localised in real and Fourier-Laguerre spaces} and their angular aperture is invariant under radial translation. We introduce a multiresolution algorithm to perform the flaglet transform rapidly, while capturing all information at each wavelet scale in the minimal number of samples on the ball.  Our implementation of these new tools achieves \bl{floating-point} precision and is made publicly available. We perform numerical experiments demonstrating the speed and accuracy of these libraries and illustrate their capabilities on a simple denoising example.
\end{abstract}


\IEEEpeerreviewmaketitle
 
\section{Introduction} 

\PARstart{A}{} common problem in data analysis is the extraction of \bl{non-trivial} patterns and structures of interest from signals. This problem can be addressed by projecting the data onto an appropriate basis. Whereas Fourier analysis focuses on oscillatory features, wavelets extract the contributions of scale-dependent features in both real and frequency space simultaneously. Initially defined in Euclidean space, wavelets have been extended to various manifolds and are now widely used in numerous disciplines. In particular, spherical wavelets \cite{antoinevander1998, antoinevander1999, wiaux2005correspondence, wiauxyvesvander2005c, starck2006ridge, mcewen2006dirwavelets, baldimarinucci2006needlets, mcewen2008dirwavelets, Marinucci2008needlets, yeo2008banks2sphere} have been extremely successful at analysing data on the sphere and have now become a standard tool in geophysics (\emph{e.g.}~\cite{audet2010topography, audet2011moongravity, simons2011, simons2011wavelets, charlety2012, loris2012}) and astrophysics (\emph{e.g.}~\cite{cobe2000gaussianityneedlets, cobe2001mexicannongaussianity, vielva2004nongaussianitywmap1, mcewen2006bianchiwmap1, starckpires2006weaklensingwavelets, vielva2006iswdarkenergy, mcewenvielva2006iswdarkenergy, mcewen2007cosmoapplications, mcewenwiaux2007darkenergy, delabrouille2008cmbestimation, marinucci2008needletsbispectrum, marinucci2008cmbaniso, starck2010poissondenoising, delabrouille2012wmap7needletilc, starckpires2012weaklensing, labatiestarck2012baos}). Naturally, data may also be defined on the three-dimensional ball when radial information (such as depth, redshift or distance, for example) is associated with each spherical map.

First approaches to perform wavelet-type transforms on the ball were developed by \cite{Michelmichel2005splinewaveletsball, Michel2005splinewaveletsball} in the continuous setting only, which thus cannot be used for exact reconstruction in practice.  The spherical Haar transform \cite{lessig2007haar, LessigFiume2008soho1} was extended to the ball by \cite{chow2010soho3dhaar} to support exact analysis and synthesis.  However, this framework is very restrictive and may not necessarily lead to a stable continuous basis \cite{schswe:sphere, swe:lift1, swe:lift2}.
The first wavelet transform on the ball to tackle both the continuous and discrete settings was developed in the influential work of Lanusse \emph{et al.}~\cite{mrs3d}. This wavelet transform is based on an isotropic undecimated wavelet construction, built on the Fourier-Bessel transform. 
Since these wavelets are isotropic, their angular aperture depends on the distance to the origin.  Although the wavelet transform on the ball is exact in Fourier-Bessel space, wavelet coefficients must be recovered on the ball from their Fourier-Bessel coefficients (in order to extract spatially localised information).  However, there exists no exact quadrature formula for the spherical Bessel transform (the radial part of the Fourier-Bessel transform) \cite{lemoine1994sbt}, and thus no way to perform the Fourier-Bessel transform exactly. Consequently, the undecimated wavelet transform on the ball is not theoretically exact when wavelet coefficients are recovered on the ball.  Nevertheless, the isotropic undecimated wavelet transform does achieve good numerical accuracy, which may be sufficient for many applications.\footnote{The accuracy of the Fourier-Bessel transform, and thus the isotropic undecimated wavelet transform on the ball, may be improved by numerical iteration, although this can prove problematic for certain applications.}
Wavelets on the ball have also been discussed in geophysics by \cite{simons2011, simons2011wavelets}, who espoused a philosophy of separability in the three Cartesian coordinates of a ball-to-``cubed-sphere-ball'' mapping, although in \cite{simons2011} examples are shown where wavelet transforms have been performed on each spherical shell only but not in the radial direction. In ongoing work, these same authors have extended their approach to the ball, where the wavelet transform in the radial direction is tailored to seismological applications by honouring certain major discontinuities in the seismic wavespeed profile of the Earth \cite{charlety2012,loris2012}.  At present, to the best of our knowledge, there does not exist an exact wavelet transform of a band-limited signal defined on the ball. 

One reason there is no exact wavelet transform on the ball is due to the absence of an exact harmonic transform.  We resolve this issue by deriving an exact spherical Laguerre transform on the radial half-line, leading to a new Fourier-Laguerre transform on the ball which is theoretically exact.  Furthermore, this gives rise to a sampling theorem on the ball, where all information of a band-limited function is captured in a finite number of samples.  With an exact harmonic transform on the ball in hand, we construct exact wavelets through a harmonic tiling, which we call \emph{flaglets} (since they are built on the Fourier-LAGuerre transform).  Each wavelet kernel is \bl{localised in real and Fourier-Laguerre spaces}, and probes a characteristic angular scale which is invariant under radial translation. Flaglets allow one to probe three-dimensional spherical data in position and scale simultaneously.  Moreover, their exactness properties guarantee that the flaglet transform captures and preserves all the information contained in a band-limited signal. 

The remainder of this article is organised as follows.  In Section \ref{harmonicanalysisontheball} we define the spherical Laguerre transform on the radial half-line and the Fourier-Laguerre transform on the ball.  In Section \ref{sec:waveletsontheball} we construct the exact flaglet transform on the ball.  In Section \ref{sec:multiresolution} we present a  multiresolution algorithm to compute the flaglet transform and evaluate our algorithms numerically. A simple denoising example is presented in Section \ref{sec:denoisingexample}. Concluding remarks are made in Section \ref{sec:conclusion}.

\section{Harmonic Analysis on the Ball} \label{harmonicanalysisontheball}

The aim of this section is to construct a novel three-dimensional transform which is appropriate for spherical coordinates and admits an exact quadrature formula. For this purpose, we first set out a radial one-dimensional transform inspired by the Laguerre polynomials and we derive a natural sampling scheme and quadrature rule on the \bl{radial half-line}. We relate this novel spherical Laguerre transform to the spherical Bessel transform \bl{and show that the latter can be evaluated exactly if the signal is band-limited in the spherical Laguerre basis}. We combine the spherical Laguerre transform with the spherical harmonics to form the Fourier-Laguerre transform on the ball, yielding a novel sampling theorem and an exact harmonic transform.\footnote{\bl{A harmonic transform is typically associated with basis functions which are eigenvalues of the Laplacian operator (e.g. the Fourier transform). In this paper our basis functions on the radial half-line (and thus on the ball) are not solutions of the Laplacian, hence harmonic analysis on the ball is interpreted in a broader sense. Nonetheless, these basis functions form orthonormal transforms and define valid dual spaces. We define band-limited signals to have bounded support in the transform space of these orthogonal basis functions.}}

\subsection{The spherical Laguerre transform}

The Laguerre polynomials, solutions to the Laguerre differential equation \cite{Weniger2008analycitylaguerre, Harry1947laguerre}, are well known for their various applications in engineering and physics, notably in the quantum-mechanical treatment of the hydrogen atom \cite{dunkl2003ortholaguerre}, as well as in modern optics \cite{Siegman73, Bond2011highordlagu}. They form a natural orthogonal basis on the interval $[0,\infty)$ (\emph{i.e.} non-negative reals $\mathbb{R}^+$) with respect to an exponential weight function. In this work, since we use this expansion along the radial half-line, we define the spherical Laguerre basis function $K_p(r)$ with $r \in \mathbb{R}^+$ as
 \begin{equation}
	K_p(r) \equiv \sqrt{ \frac{p!}{(p+2)!} }  \frac{ e^{-{r}/{2\tau}} }{ \sqrt{\tau^3}} L^{(2)}_p\left(\frac{r}{\tau}\right),
\end{equation}
where $L^{(2)}_p$ is the $p$-th generalised Laguerre polynomial of order two, \bl{defined as
\begin{equation}
	L_p^{(2)}(r) \equiv \sum_{j=0}^{p}  {p+2 \choose p - j} \frac{(-r)^j}{j!}, \label{lagudef}
\end{equation}
}%
and $\tau \in \mathbb{R}^+$ is a scale factor that adds a scaling flexibility and shall be defined at the end of this section. The basis functions \bl{$K_p$} are orthonormal on $\mathbb{R}^+$ with respect to a radial inner product:
\begin{equation}
	 \langle K_p | K_q \rangle = \int_{\mathbb{R}^+} {\rm d} r r^2 K_p(r) K^*_q(r) = \delta_{pq}. \label{orthobasis}
\end{equation}
Note that the complex conjugate ${}^*$ is facultative since we use real basis functions. Any square-integrable real signal \mbox{$f\in L^2(\mathbb{R}^+)$} may be expanded as
\begin{equation}
	f(r) = \sum_{p=0}^\infty {f}_p K_p(r) \label{laguerrebackward},
\end{equation}
for natural $p\in\mathbb{N}$, where ${f}_p$ is the projection of $f$ onto the $p$-th basis function:
\begin{equation}
	{f}_p = \langle f | K_p \rangle = \int_{\mathbb{R}^+} {\rm d} r r^2 f(r) K^*_p(r). \label{laguerreanalysis}
\end{equation}
The decomposition follows by the orthogonality and completeness of the spherical Laguerre basis functions: orthonormality is given by Eqn.~(\ref{orthobasis}), while the completeness relation is obtained by applying the Gram-Schmidt orthogonalisation process to the basis functions and exploiting the completeness of polynomials on $L^2(\mathbb{R}^+, r^2 e^{-r}{\rm d} r)$.

\begin{figure}\centering
\includegraphics[trim=9cm 1cm 2cm 0.5cm, clip, width=8.5cm]{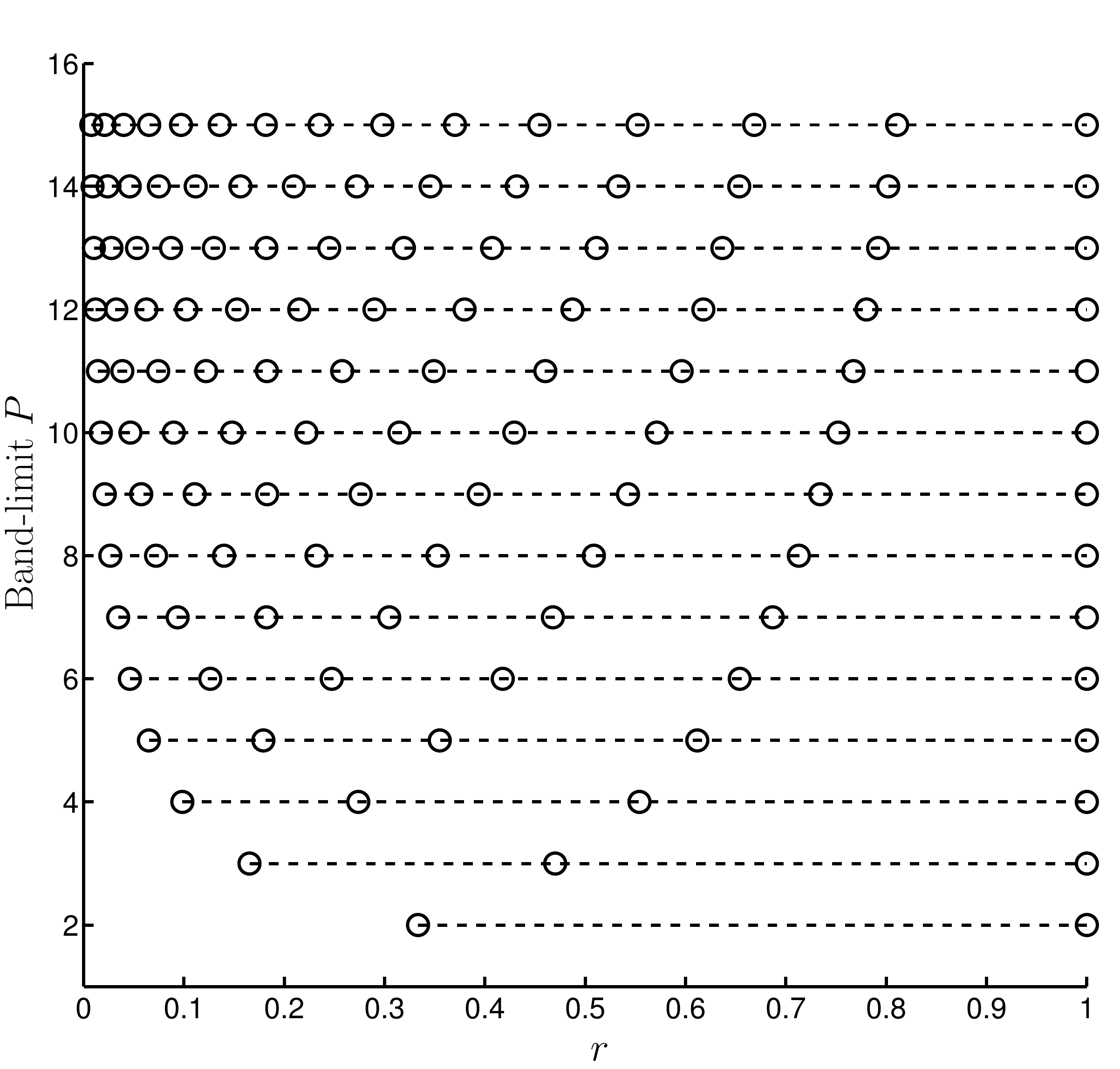}
\caption{Spherical Laguerre sampling scheme on $r \in [0,1]$ for increasing band-limit $P$. If a function $f$ is $P$-band-limited then $f$ and the basis functions need only be evaluated on $P$ points for the spherical Laguerre transform to be exact. For a particular $P$, the associated sampling is denser near the origin \bl{since the quadrature is constructed on $\mathbb{R}^+$ with measure $e^{-r}dr$.}}
\label{fig:rootslaguerre}
\end{figure}

\begin{figure}\centering
\subfigure[Basis functions {$K_p(r)$}]{\includegraphics[trim=7.5cm 23.2cm 1.5cm 0cm, clip, width=8cm]{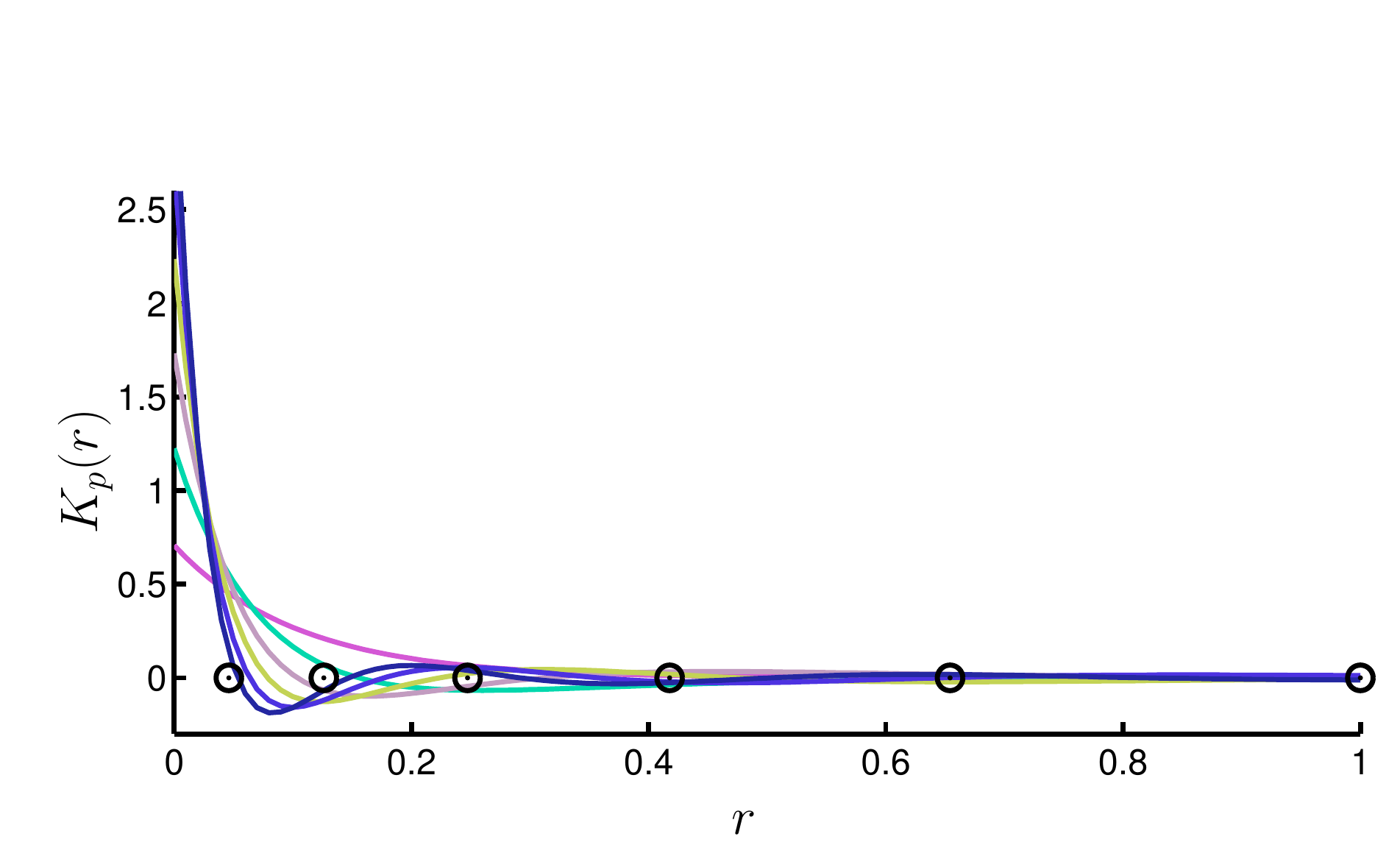}}\\
\subfigure[Zoom on the oscillatory features of {$K_p(r)$}]{\includegraphics[trim=7.5cm 12.7cm 1.5cm 12cm, clip, width=8cm]{pics/sphelaguerrebasis.pdf}}\\
\subfigure[Functions $rK_p(r)$]{\includegraphics[trim=7.5cm 2.3cm 1.5cm 23cm, clip, width=8cm]{pics/sphelaguerrebasis.pdf}}
\caption{First six spherical Laguerre basis functions $K_p(r)$ constructed on $r \in [0,1]$ and the associated sample positions (circles). A function $f$ with band-limit $P=6$ can be decomposed and reconstructed exactly using these six basis functions only. In that case, $f$ and the basis functions are solely evaluated at the sampling points. Functions $rK_p(r)$ can be viewed as basis functions in cartesian coordinates satisfying the usual orthogonality relation $\int_{\mathbb{R}^+} {\rm d} r (r K_p(r) )(rK_q(r)) = \delta_{pq}$.}
\label{fig:spherlaguerrebasis}
\end{figure}

When it comes to calculating the transform, one must evaluate the integral of Eqn.~(\ref{laguerreanalysis}) numerically.   We consider functions $f$ band-limited at $P$ \bl{in the spherical Laguerre basis}, such that $f_p = 0, \ \forall p\geq P$. It is straightforward to show that if $f$ is band-limited, then both \bl{$e^{r/2\tau}K_p(r)$ and $e^{r/2\tau}f(r)$} are polynomials of maximum degree $P-1$. In this case, Eqn.~(\ref{laguerreanalysis}) is the integral of a polynomial of order $2P-2$ on $\mathbb{R}^+$ with weight function $r^2 e^{-r}$.  Thus, applying Gaussian quadrature (\emph{e.g.} \cite{numericalrecipes, gauss2012}) with $P$ sampling nodes is sufficient to evaluate this integral exactly. The resulting quadrature formula is known as the Gauss-Laguerre quadrature and is commonly used to evaluate numerical integrals on $\mathbb{R}^+$. Hence, Eqn.~(\ref{laguerreanalysis}) \bl{reduces to a} weighted sum:
\begin{equation}
	{f}_p = \sum_{i=0}^{{P-1}} w_i  f(r_i) K^*_p(r_i) \label{gausslaguquadr},
\end{equation}
where $r_i\in\mathbb{R}^+$ is the $i$-th root of the $P$-th generalised Laguerre polynomial of order two, and 
\bl{\begin{equation}
  w_i =  \frac{(P+2)r_i e^{r_i}}{(P+1)[  L^{(2)}_{P+1}(r_i)]^{2}}\in\mathbb{R}^+
\end{equation}}%
is the corresponding weight. \bl{Any $P$-band-limited function $f$ can be decomposed and reconstructed exactly using the spherical Laguerre transform. All information content of the function is captured in $P$ samples located in the interval $[0,r_{P-1}]$ where $r_{P-1}$ is the largest root of the sampling. Since $r_{P-1}$ increases with $P$, one may wish to rescale the sampling so that the spherical Laguerre transform contains samples in any interval of interest $[0,R]$, with $R\in\mathbb{R}^+$, while the underlying continuous function is nevertheless defined on $\mathbb{R}^+$. The scale factor $\tau$ is then chosen such that $\tau = R/r_{P-1}$. Figure~\ref{fig:rootslaguerre} shows the resulting spherical Laguerre sampling constructed on $r \in [0,1]$ (\emph{i.e.} rescaled with $\tau$) for increasing band-limit $P$. Figure~\ref{fig:spherlaguerrebasis} shows the first six basis functions constructed on $r \in [0,1]$ and the sampling nodes used to obtain an exact transform.\footnote{If one preferred to consider the measure ${\rm d}r$ rather than the spherical measure $r^2 {\rm d} r $, then the basis functions $r K_p(r)$ shown in Figure~\ref{fig:spherlaguerrebasis}~(c) could be used in place of the spherical Laguerre basis functions defined here.} Note that the spherical Laguerre transform is a real transform that can be extended to complex signals by considering the real and imaginary parts separately.}

\subsection{Relation to the spherical Bessel transform}\label{sec:spherbessel}

The spherical Bessel transform is a fundamental radial transform arising from the resolution of the Laplacian operator in spherical coordinates. It is central to the Fourier-Bessel transform, commonly used in cosmology \cite{leistedt20113dex, rassat2011baos, abramo2010cmbbox} to analyse the spectral properties of galaxy surveys in three dimensions. In this section we derive an analytical formula to exactly compute the spherical Bessel transform of a function whose spherical Laguerre transform is band-limited. This section is optional to the reader interested in wavelets only.

The spherical Bessel transform of $f \in L^2(\mathbb{R}^+)$ reads
\bl{\begin{equation}
	\tilde{f}_\ell(k) =   \langle f | j_\ell \rangle = \sqrt{\frac{2}{\pi}} \int_{\mathbb{R}^+} {\rm d} r r^2 f(r) j_\ell^*(kr), \label{besselforward}
\end{equation}
for $k \in \mathbb{R}^+$, $\ell \in \mathbb{N}$, and where $j_\ell(kr)$ is the $\ell$-th order spherical Bessel function. Note again that the complex conjugate is facultative since the spherical Bessel functions are real.} The reconstruction formula is given by
 \begin{equation}
	f(r) =  \sqrt{\frac{2}{\pi}} \int_{\mathbb{R}^+} {\rm d}k k^2 \tilde{f}_\ell(k) j_\ell(kr).
\end{equation}
The spherical Bessel transform is thus symmetric and the problem is reduced to the calculation of a similar inner product for the decomposition and the reconstruction. However, to our knowledge, there exists no method to compute such an integral exactly \bl{for a useful class of functions}, and finding a quadrature formula for the spherical Bessel functions on $\mathbb{R}^+$ is a non-trivial issue. Moreover, the use of numerical integration methods does not always guarantee good accuracy because of the oscillatory nature of the spherical Bessel functions.

To find a tractable expression to compute Eqn.~(\ref{besselforward}), we first express $f$ by its spherical Laguerre expansion, giving
\begin{equation}
	\tilde{f}_\ell(k) = \sqrt{\frac{2}{\pi}} \sum_p {f}_p {j}_{\ell p} (k)  \label{exactbessel},
\end{equation}
which is a finite sum if $f$ is band-limited in spherical Laguerre space. In this expression ${j}_{\ell p} (k)$ is the projection of $K_p$ onto $j_\ell(kr)$, i.e.
 \begin{equation}
 	{j}_{\ell p} (k) \equiv \langle K_p | j_\ell \rangle =   \int_{\mathbb{R}^+} {\rm d} r r^2 K_p(r)  j^*_\ell(k r).\label{bessellaguerreproblem}
 \end{equation}
Consequently, the problem of computing the spherical Bessel decomposition of $f$ is recast as evaluating Eqn.~(\ref{exactbessel}), through the computation of the inner product of Eqn.~(\ref{bessellaguerreproblem}). But unlike the initial problem of Eqn.~(\ref{besselforward}), ${j}_{\ell p} (k)$ admits an analytic formula. Starting from the definition of Laguerre polynomials \bl{in Eqn.~(\ref{lagudef})}, one can show that 
\begin{equation}
	{j}_{\ell p} (k)  =  \sqrt{ \frac{p!}{(p+2)!} }  \sum_{j=0}^{p}   c^{p}_{j} \mu^\ell_{j+2} (k), \label{jlpk} 
\end{equation}
where the $c^p_{j}$ satisfy the following recurrence:\bl{
\begin{equation}
	c^{p}_{j} \equiv \frac{(-1)^j}{j!} {p+2 \choose p - j} =  -\frac{p-j+1}{j(j+2)} c^{p}_{j-1}.
\end{equation}}%
The functions $\mu^\ell_{j} (k)$ are the moments of $j_\ell(kr)e^{-\frac{r}{2\tau}}$, \emph{i.e.} 
\begin{equation}
	\mu^\ell_{j} (k) \equiv \frac{1}{\tau^{j-\frac{1}{2}}} \int_{\mathbb{R}^+} {\rm d} r r^j j_\ell(kr) e^{-\frac{r}{2\tau}}.
\end{equation}
From \cite{watson1995treatise} we find an analytical solution for the latter integral:
\begin{eqnarray}\small
	\quad \mu^\ell_j(k)   &=& \ \sqrt{\pi} \ 2^{j}\ \tilde{k}^{\ell} \ \tau^\frac{3}{2}  \ \frac{  \Gamma(j + \ell + 1) }{ \Gamma(\ell+\frac{3}{2}) }  \label{muljk} \\ 
	& \times & \hspace{-2mm} \phantom{s}_2F_1 \left( \frac{j + \ell + 1}{2}  ;  \frac{ j + \ell}{2} + 1 ; \ell+\frac{3}{2} ; -4\tilde{k}^2 \right) \nonumber
\end{eqnarray}
where $\tilde{k} = \tau k$ is the rescaled $k$ scale and$\phantom{s}_2F_1$ is the Gaussian hypergeometric function. \bl{Since either $(j+\ell+1)/2$ or $(j+\ell)/2$ is a positive integer, the latter reduces to a polynomial of $\tilde{k}^2$ and it is possible to compute the quantity ${j}_{\ell p} (k)$ exactly using Eqn.~(\ref{jlpk}) to (\ref{muljk}).} Consequently, the inverse spherical Bessel transform $\tilde{f}_\ell(k)$ may then be calculated analytically through Eqn.~(\ref{exactbessel}), which is computed exactly if $f$ is band-limited \bl{in the spherical Laguerre basis.}

\subsection{The spherical harmonic transform}

Whereas the spherical Laguerre transform is specifically designed for analysing functions on the \bl{radial half-line}, the spherical harmonic transform is a natural choice for the angular part of a consistent three-dimensional analysis. For a function \bl{$f \in L^2(S^2)$} on the two-dimensional sphere, the transform reads 
\begin{equation}
	f(\omega) = \sum_{\ell = 0}^\infty \sum_{m = -\ell}^\ell f_{\ell m} Y_{\ell m}(\omega),
\end{equation}
where $\omega=(\theta, \phi)\in S^2$ are spherical coordinates of the unit sphere $S^2$, with colatitude $\theta \in [0,\pi]$ and longitude $\phi \in [0,2\pi)$.  Thanks to the orthogonality and completeness of the spherical harmonics $Y_{\ell m}(\omega)$, the inverse transform is given by the following inner product on the sphere:
\begin{equation}
	f_{\ell m} = \langle f | Y_{\ell m} \rangle = \int_{S^2} {\rm d}\omega f(\omega) Y^*_{\ell m}(\omega), \label{harmonicsanalysis}
\end{equation}
with surface element ${\rm d}\omega = \sin\theta {\rm d}\theta {\rm d}\phi$. For a function which is band-limited in this basis at $L$, \emph{i.e.} $f_{\ell m} = 0, \ \forall \ell \geq L$, the decomposition and reconstruction operations can be performed with a finite summation over the harmonics. This is usually resolved by defining an appropriate sampling theorem on the sphere with nodes $\omega_j=(\theta_j, \phi_j)$, associated with a quadrature formula. Various sampling theorems exist in the literature \cite{driscollhealy1994, Healy96fftsfor, mcewen2011novelsampling}; the main features of all sampling theorems are (i) the number of nodes required to capture all information in a band-limited signal and (ii) the complexity of the related algorithms to compute forward and inverse spherical harmonic transforms. Although this work is independent from this choice (provided that it leads to an exact transform), we adopt the McEwen \& Wiaux (hereafter MW) sampling theorem \cite{mcewen2011novelsampling} which is equiangular and has the lowest number of samples for a given band-limit $L$, namely $(L-1)(2L-1)+1\sim2L^2$. The corresponding algorithms to compute the spherical harmonic transforms scale as $\mathcal{O}(L^3)$ and are numerically stable to band-limits of at least $L = 4096$ \cite{mcewen2011novelsampling}.
Further technical details are provided in Section \ref{sec:implementation}. 

\subsection{The Fourier-Laguerre transform}    

We define the Fourier-Laguerre basis functions on \mbox{\bl{$B^3= {\mathbb{R}^+} \times S^2$}} as the product of the spherical Laguerre basis functions and the spherical harmonics: \mbox{$Z_{\ell m p}(\vect{r}) = K_p(r) Y_{\ell m}(\omega)$} with the 3D spherical coordinates \mbox{$\vect{r} = (r, \omega) \in B^3 $}.  The orthogonality and completeness of the Fourier-Laguerre basis functions follow from the corresponding properties of the individual basis functions, where the orthogonality relation is given explicitly by the following inner product on \bl{$B^3$}:
\begin{eqnarray}
	 \langle Z_{\ell m p} | Z_{{\ell^\prime} m^\prime p^\prime} \rangle 
         &=& \int_{B^3} {\rm d}^3\vect{r} Z_{\ell m p} Z^*_{\ell^\prime m^\prime p^\prime} (\vect{r}) \\
         &=& \delta_{\ell \ell^\prime} \delta_{mm^\prime} \delta_{pp^\prime}, \nonumber
\end{eqnarray}
where ${\rm d}^3\vect{r} = r^2 \sin\theta {\rm d}r {\rm d}\theta {\rm d}\phi$ is the volume element in spherical coordinates. Any three-dimensional signal $f \in L^2(B^3)$ can be decomposed as
\begin{equation}
	f(\vect{r}) = \sum_{p = 0}^{P-1}\sum_{\ell = 0}^{L-1}\sum_{m = -\ell}^{\ell} f_{\ell m p} Z_{\ell m p} (\vect{r}), \label{flagforward}
\end{equation}
with $L$ and $P$ the angular and radial band-limits, respectively, \emph{i.e.} $f$ is such that \mbox{$f_{\ell m p} = 0$}, $\forall \ell \geq L$,  $\forall p \geq P$. The inverse relation is given by the projection of $f$ onto the basis functions:
\begin{equation}
	{f}_{\ell m p} = \langle f | Z_{\ell m p} \rangle =   \int_{B^3} {\rm d}^3\vect{r}  f(\vect{r}) Z^*_{\ell m p}(\vect{r}).  \label{flaginverse}
\end{equation}
The Fourier-Laguerre transform may also be related to the \bl{Fourier-Bessel transform using the results of Section \ref{sec:spherbessel}}.\footnote{The Fourier-Laguerre and the Fourier-Bessel transforms of $f$ are related through
\begin{equation*} 
	\tilde{f}_{\ell m}(k) = \sqrt{\frac{2}{\pi}} \sum_p {f}_{\ell m p}  {j}_{\ell p}(k). 
\end{equation*}
If $f$ is band-limited in terms of its Fourier-Laguerre decomposition, the latter sum is finite and both transforms can be calculated exactly since ${j}_{\ell p}(k)$ admit the exact analytic formula Eqn.~(\ref{jlpk}).
}

\bl{In practice, calculating the Fourier-Laguerre transform requires the evaluation of the integral of Eqn.~(\ref{flaginverse}). For this purpose, combining the quadrature rules on the sphere and on the radial half-line leads to a sampling theorem on $B^3$.  For a signal with angular and radial band-limits $L$ and $P$, respectively, all of the information content of the signal is captured in $N = P[(2L-1)(L-1)+1] \sim 2PL^2$ samples, yielding an exact Fourier-Laguerre transform on $B^3$. The three-dimensional sampling consists of spherical shells, discretised according to a sampling theorem (where here we adopt the MW sampling theorem), located at the nodes of the radial sampling. The radial sampling may furthermore be rescaled to any spherical region of interest $[0,R] \times S^2$ using the parameter $\tau$ to dilate or contract the radial quadrature rule.}

\section{Wavelets on the Ball} \label{sec:waveletsontheball}

The exactness of the Fourier-Laguerre transform supports the design of an exact wavelet transform on the ball. In this section we first define a three-dimensional convolution operator on the ball, derived from the  convolutions defined on the sphere and on the radial half-line. We then construct flaglets through an exact tiling of \bl{Fourier-Laguerre} space, leading to wavelet kernels which are \bl{spatially localised and form a tight frame}.  Furthermore, each kernel projects onto an angular scale which is invariant under radial translation. We finally introduce a multiresolution algorithm to compute the flaglet transform and capture the information of each wavelet scale in the minimal number of samples on the ball, while optimising the computational cost of the transform.

\subsection{Convolutions} 

The convolution of two functions $f$ and $h$ in a (Hilbert) space of interest is often defined by the inner product of $f$ with a transformed version of $h$.  In standard Fourier analysis this transformation is the natural translation. Likewise, for two signals on the sphere $f, h\in L^2(S^2)$, the convolution is constructed from the rotation operator $\mathcal{R}_\omega$: 
\begin{equation}
	(f \star h)(\omega) \equiv \langle f | \mathcal{R}_\omega h \rangle = \int_{S^2} {\rm d}\omega^\prime f(\omega^\prime) \left( \mathcal{R}_\omega h \right)^*(\omega^\prime).
\end{equation}
where, here and henceforth, we restrict ourselves to axisymmetric kernels $h$, so that the rotation is only parameterised by an angle $\omega=(\theta,\phi)$ \cite{mcewen2007dirwavelets, mcewen2008dirwavelets}.  The spherical harmonic decomposition of $f \star h$ is given by the product of the individual transforms:
\begin{equation}
	{(f \star h)}_{\ell m} = \langle f \star h | Y_{\ell m} \rangle =  \sqrt{ \frac{4\pi}{2\ell+1}} {f}_{\ell m} {h}^*_{\ell 0},
\end{equation}
with $f_{\ell m} = \langle f|Y_{\ell m} \rangle$ and $h_{\ell 0}\delta_{m 0} = \langle h|Y_{\ell m} \rangle$. 

\begin{figure}\centering
\subfigure[Wavelet kernel translated by $r=0.2$]{\includegraphics[trim= 1cm 24.4cm 0.5cm 1cm, clip, width=4cm]{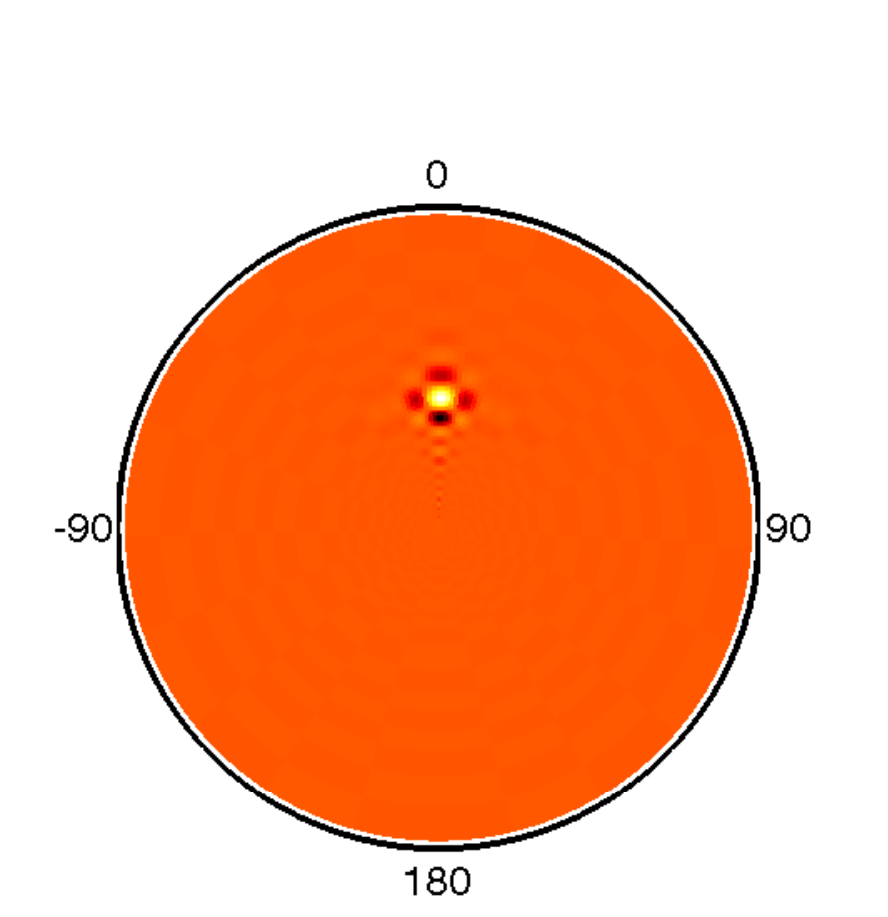}\includegraphics[trim= 6cm 23.4cm 1cm 1cm, clip, width=4cm]{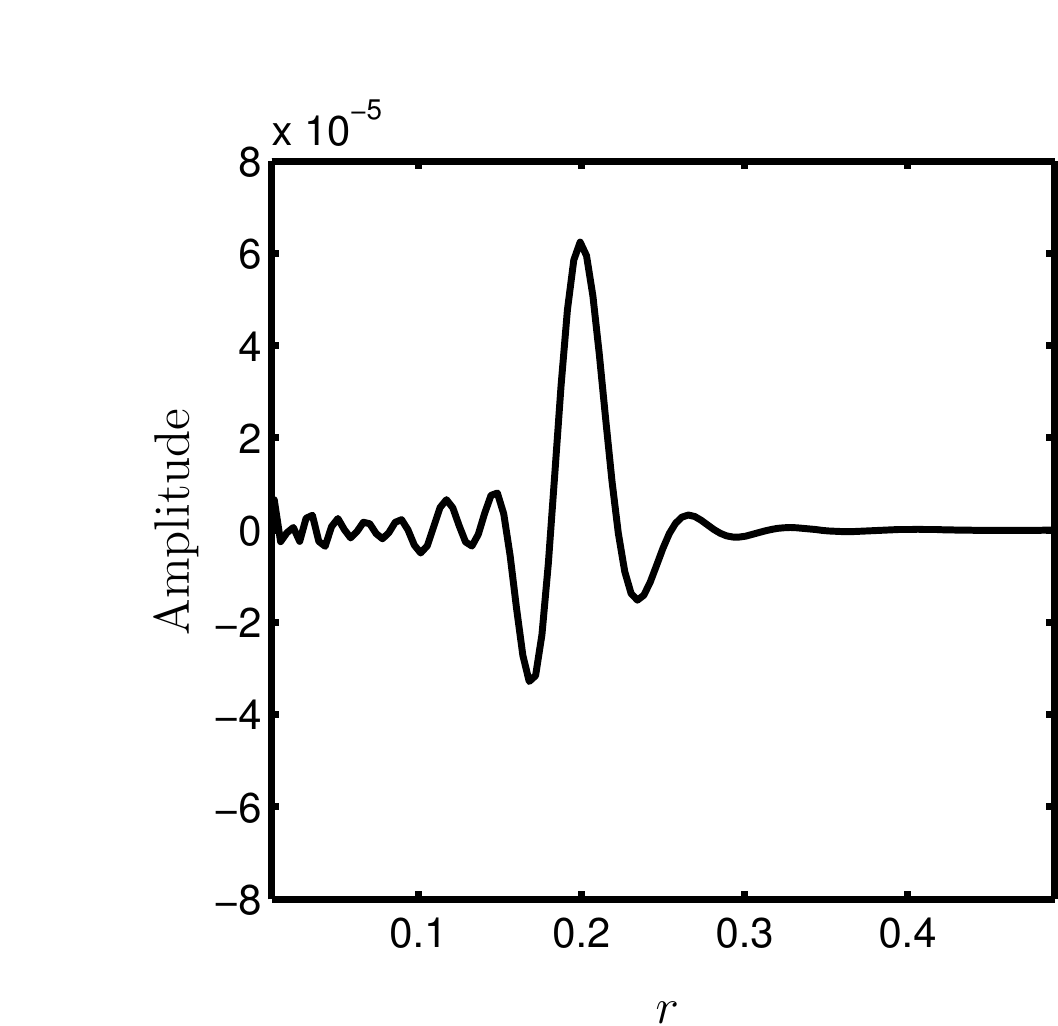}}\\
\subfigure[Wavelet kernel translated by $r=0.3$]{\includegraphics[trim= 1cm 14cm 0.5cm 12.5cm, clip, width=4cm]{pics/translationoperator_bis.pdf}\includegraphics[trim= 6cm 12.9cm 1cm 12.4cm, clip, width=4cm]{pics/translationoperator.pdf}}\\
\subfigure[Wavelet kernel translated by $r=0.4$]{\includegraphics[trim= 1cm 3.6cm 0.5cm 22.8cm, clip, width=4cm]{pics/translationoperator_bis.pdf}\includegraphics[trim= 6cm 2.6cm 1cm 22.8cm, clip, width=4cm]{pics/translationoperator.pdf}}
\caption{Slices of an axisymmetric flaglet wavelet kernel constructed on the ball of radius $R=1$, translated along the radial half-line. The chosen kernel has $j=j^\prime=5$ and is constructed at resolution $P=L=64$. For clarity we zoomed on the range $r\in[0,0.5]$ (the slice hence relates to a ball of radius $r=0.5$). The three-dimensional wavelet can be visualised by rotating this slice around the vertical axis passing through the origin. The translation on the radial half-line not only translates the main feature (the wavelet peak) but also accounts for a damping factor. Flaglets are well localised in both real and \bl{Fourier-Laguerre} spaces and their angular aperture is invariant under radial translation.} 
\label{fig:wavtransl}
\end{figure}

Similarly, we introduce a translation operator $\mathcal{T}_r$ to construct the convolution of two functions on the radial half-line \mbox{$f, h \in L^2(\mathbb{R}^+)$}:
\begin{equation}
	(f \star g)(r) \equiv \langle f | \mathcal{T}_r h \rangle = \int_{\mathbb{R}^+} {\rm d} r^\prime r^{\prime2} f(r^\prime) \left( \mathcal{T}_r h \right)^*(r^\prime).
\end{equation}
The convolution in Laguerre space \cite{Gorlich1982laguerreconvolution, markett1986, Kanjin1986} is defined such that the action $\mathcal{T}_r$ on the basis functions is
\begin{equation}
	(\mathcal{T}_r K_p )(r^\prime) \equiv K^*_p(r) K_p(r^\prime), \label{laguprodconv}
\end{equation}
in which case $f \star h$ simplifies to a product in spherical Laguerre space, yielding
\begin{equation}
	{(f \star h)}_p  = \langle f \star h | K_p \rangle = {f}_p {h}_p^*,
\end{equation}
where $f_p = \langle f|K_p \rangle$ and $h_p = \langle h|K_p \rangle$. Consequently any function $f$ which is translated by a distance $r$ on the radial half-line has each coefficient $f_p$ transformed into $f_p K_p(r)$. 
This operation corresponds to a translation with a damping factor, which is illustrated on a wavelet kernel in Figure~\ref{fig:wavtransl} (the wavelet kernel itself is defined in Section~\ref{sec:waveletkernels}).\footnote{Note that this translation operator may also be viewed in real space as a convolution with a delta function, similarly to the Euclidian convolution.}

Finally, we define the convolution of two functions on the ball $f, h \in L^2(B^3)$, where $h$ is again assumed to be axisymmetric in the angular direction, by combining the convolution operators defined on the sphere and radial half-line, yielding
\begin{eqnarray}
	(f \star h)(\vect{r}) &\equiv& \langle f | \mathcal{T}_r  \mathcal{R}_\omega h \rangle  \\
	&= &\int_{B^3} {\rm d}^3\vect{r}^\prime f(\vect{r}^\prime) \left( \mathcal{T}_r \mathcal{R}_\omega h \right)^*(\vect{r}^\prime).
\end{eqnarray}
The convolution is given in harmonic space by the product
\begin{equation}
	{(f \star h)}_{\ell m p}=  \langle f \star h | Z_{\ell m p} \rangle =  \sqrt{ \frac{4\pi}{2\ell+1}} {f}_{\ell mp} {h}^*_{\ell 0 p} ,
\end{equation}
with $f_{\ell m p} = \langle f|Z_{\ell m p} \rangle$ and $h_{\ell 0 p}\delta_{m 0} = \langle h|Z_{\ell mp} \rangle$.

\subsection{Exact flaglet transform}

With an exact harmonic transform and a convolution operator defined on the ball in hand, we are now in a position to construct the exact flaglet transform on the ball. For a function of interest $ f\in L^2(B^3)$, we define its $jj^\prime$-th wavelet coefficient $W^{\Psi^{jj^\prime}}\in L^2(B^3)$ as the convolution of $f$ with the flaglet (\emph{i.e.} wavelet kernel) $\Psi^{jj^\prime}\in L^2(B^3)$:
\begin{equation}
 	W^{\Psi^{jj^\prime}}(\vect{r})  \equiv (f \star \Psi^{jj^\prime})(\vect{r}) = \langle f | \mathcal{T}_r \mathcal{R}_\omega \Psi^{jj^\prime} \rangle. \label{wav1}
\end{equation} 
The scales $j$ and $j^\prime$ respectively relate to angular and radial spaces. Since we restrict ourselves to axisymmetric kernels, the wavelet coefficients are given in \bl{Fourier-Laguerre} space by the product
\begin{equation}
	{W}^{\Psi^{jj^\prime}}_{\ell m p} =  \sqrt{ \frac{4\pi}{2\ell+1}} {f}_{\ell m p} {\Psi}^{jj^\prime*}_{\ell 0 p}, \label{wav2}
\end{equation}
where ${W}^{\Psi^{jj^\prime}}_{\ell m p} = \langle W^{\Psi^{jj^\prime}} |Z_{\ell m p} \rangle$, $f_{\ell m p} = \langle f|Z_{\ell m p} \rangle$ and ${\Psi}^{jj^\prime}_{\ell 0 p}\delta_{m 0} = \langle \Psi^{jj^\prime}|Z_{\ell m p} \rangle$.  The wavelet coefficients contain the detail information of the signal only; a scaling function and corresponding scaling coefficients must be introduced to represent the low-frequency, approximate information of the signal.  The scaling coefficients $W^\Phi \in L^2(B^3)$ are defined by the convolution of $f$ with the scaling function $\Phi\in L^2(B^3)$:
\begin{equation}
W^\Phi(\vect{r})  \equiv (f \star \Phi)(\vect{r}) = \langle f | \mathcal{T}_r \mathcal{R}_\omega \Phi \rangle, \label{wav3}
\end{equation}
or in \bl{Fourier-Laguerre} space,
\begin{equation}
	{W}^\Phi_{\ell m p} =  \sqrt{ \frac{4\pi}{2\ell+1}} {f}_{\ell m p} {\Phi}^{*}_{\ell 0 p}, \label{wav4}
\end{equation}
where ${W}^{\Phi}_{\ell m p} = \langle W^{\Phi} |Z_{\ell m p} \rangle$ and ${\Phi}_{\ell 0 p}\delta_{m 0} = \langle \Phi|Z_{\ell m p} \rangle$. 

Provided the flaglets and scaling function satisfy an admissibility property, a function $f$ may be reconstructed exactly from its wavelet and scaling coefficients by
\begin{eqnarray}
	\quad f(\vect{r}) \ \ = \ \ \int_{B^3} {\rm d}^3\vect{r}^\prime W^{\Phi}(\vect{r}^\prime)(\mathcal{T}_r \mathcal{R}_\omega \Phi)(\vect{r}^\prime) \ \ \ \ \ \ \  \\
	+ \ \sum_{j=J_0}^{J} \sum_{j^\prime=J^\prime_0}^{J^\prime}  \int_{B^3} {\rm d}^3\vect{r}^\prime W^{\Psi^{jj^\prime}}\hspace{-2mm}(\vect{r}^\prime)(\mathcal{T}_r \mathcal{R}_\omega \Psi^{jj^\prime})(\vect{r}^\prime), \nonumber
\end{eqnarray}
or equivalently in harmonic space by
\begin{eqnarray}
	 {f}_{\ell m p} &=& \sqrt{ \frac{4\pi}{2\ell+1}} {W}^{\Phi}_{\ell m p} {\Phi}_{\ell 0 p}  \\ 
	&+& \ \sqrt{ \frac{4\pi}{2\ell+1}} \sum_{j=J_0}^{J} \sum_{j^\prime=J^\prime_0}^{J^\prime}   {W}^{\Psi^{jj^\prime}}_{\ell m p} {\Psi}^{jj^\prime}_{\ell 0 p}. \label{waveletsynthesis}\nonumber
\end{eqnarray}
The parameters $J_0$, $J^\prime_0$, $J$ and $J^\prime$ defining the minimum and maximum scales must be defined consistently to extract and reconstruct all the information contained in $f$. They depend on the construction of the flaglets and scaling function and are defined explicitly in the next section. 

Finally, the admissibility condition under which a band-limited function $f$ can be decomposed and reconstructed exactly is given by the following resolution of the identity:
\begin{equation}
	\frac{4\pi}{2\ell+1} \left( |{\Phi}_{\ell 0 p}|^2 + \sum_{j=J_0}^{J} \sum_{j^\prime=J^\prime_0}^{J^\prime}  |{\Psi}^{jj^\prime}_{\ell 0 p}|^2 \right) \ = \ 1, \quad \forall \ell, p . \label{identity}
\end{equation}
We may now construct flaglets and scaling functions that satisfy this admissibility property and thus lead to an exact wavelet transform on the ball. 

\subsection{Flaglets and scaling functions}
\label{sec:waveletkernels}

We extend the notion of harmonic tiling \cite{Marinucci2008needlets, mcewen2008dirwavelets, 2010needatool} to the Fourier-Laguerre space and construct axisymmetric wavelets (flaglets) \bl{well localised in both real and \bl{Fourier-Laguerre} spaces}.  We first define the flaglet and scaling function generating functions, before defining the flaglets and scaling function themselves.

We start by considering \bl{the $C^{\infty}$ Schwartz function with compact support}
\begin{equation}
	s(t) \equiv \left\{ \begin{array}{ll} \ e^{-\frac{1}{1-t^2}}, & t\in[-1,1] \\ \  0, & t \notin [-1,1]\end{array} \right. ,
\end{equation}
for $t \in \mathbb{R}$.  We introduce the positive real parameter $\lambda\in\mathbb{R}^+_*$ to map $s(t)$ to 
\begin{equation}
	s_\lambda(t) \equiv s\left( \frac{2\lambda}{\lambda-1} (t-1/\lambda)-1\right),
\end{equation}
which has compact support in $[\frac{1}{\lambda}, 1]$.  We then define the smoothly decreasing function $k_\lambda$ by
\begin{equation}
	 k_\lambda(t) \equiv \frac{\int_{t}^1\frac{{\rm d}t^\prime}{t^\prime}s_\lambda^2(t^\prime)}{\int_{1/\lambda}^1\frac{{\rm d}t^\prime}{t^\prime}s_\lambda^2(t^\prime)}, \label{smoothscaling}
\end{equation}
which is unity for $t<1/\lambda$, zero for $t>1$, and is smoothly decreasing from unity to zero for $t \in [1/\lambda,1]$. 
Axisymmetric flaglets are constructed in a two-dimensional space corresponding to the harmonic indices $\ell$ and $p$. We associate $\lambda$ with $\ell$-space and we introduce a second parameter $\nu$ associated with $p$-space, with the corresponding functions $s_\nu$ and $k_\nu$. We define the flaglet generating function by
\begin{equation}
	 \kappa_\lambda(t) \equiv \sqrt{ k_\lambda(t/\lambda) - k_\lambda(t) }
\end{equation}
and the scaling function generating function by
\begin{equation}
	 \eta_{\lambda}(t) \equiv \sqrt{ k_\lambda(t)} ,
\end{equation}
with similar expressions for \bl{$\kappa_\nu$ and $\eta_\nu$}, complemented with a hybrid scaling function generating function
\begin{eqnarray}
	\eta_{\lambda \nu}(t, t^\prime) &\equiv & \left[ \ \ \ k_\lambda(t/\lambda)k_\nu(t^\prime) \right. \nonumber \\
		&& \ + \ k_\lambda(t)k_\nu(t^\prime/\nu)  \\
		&& \ - \ \left. k_\lambda(t)k_\nu(t^\prime) \ \ \ \right]^{1/2} \nonumber.
\end{eqnarray}

\begin{figure}\centering
\setlength{\unitlength}{.5in}
\begin{picture}(9,7)(0,0)
\put(0.9,0.){\includegraphics[trim = 10.9cm 0cm 1cm 0cm, width=8cm]{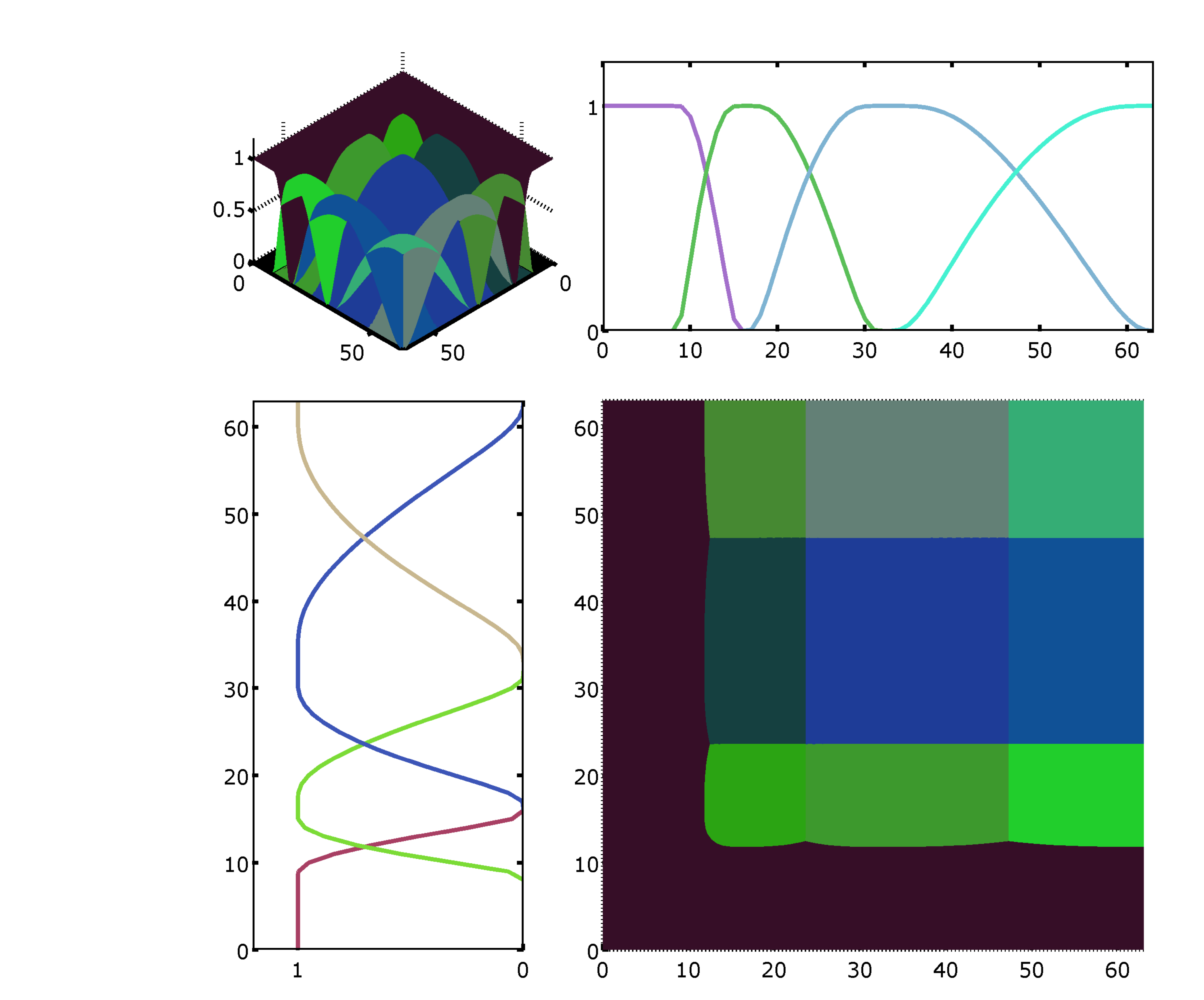}}
\put(0.7,4.7){\scriptsize$\ell$}
\put(2.3,4.7){\scriptsize$p$}
\put(0.13,2.2){\scriptsize$\ell$}
\put(2.65,2.2){\scriptsize$\ell$}
\put(4.87,0){\scriptsize$p$}
\put(4.87,4.52){\scriptsize$p$}
\end{picture}
\caption{Tiling of Fourier-Laguerre space at resolution $L=N=64$ for flaglet parameters $\lambda=\nu=2$, giving $J=J^\prime=7$. Flaglets divide \bl{Fourier-Laguerre} space into regions corresponding to specific scales in angular and radial space. The scaling part, here chosen as $J_0=J^\prime_0=4$, is introduced to cover the low frequency region and insures that large scales are also represented by the transform.} 
\label{fig:tiling}
\end{figure}
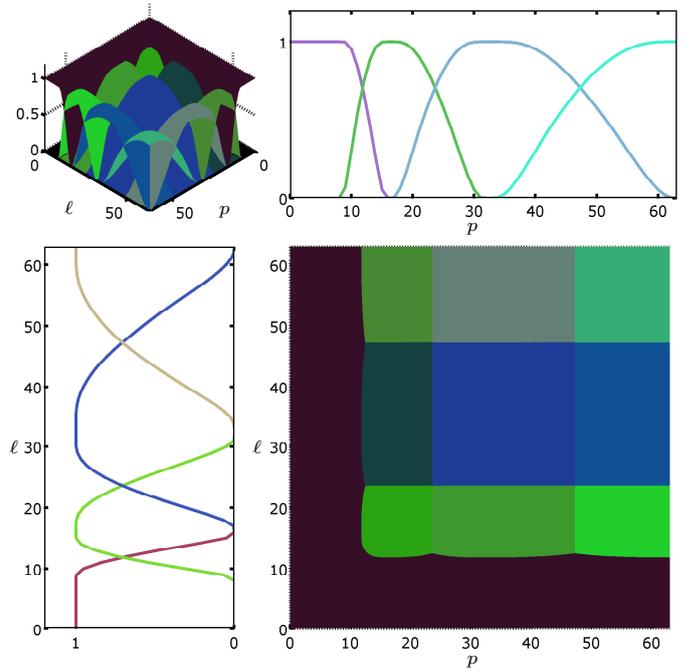

\begin{figure}\centering
\subfigure[$(j, j^\prime) = (4,5)$]{\includegraphics[trim = 10.cm 16.5cm 13.3cm 2cm, clip, width=4cm]{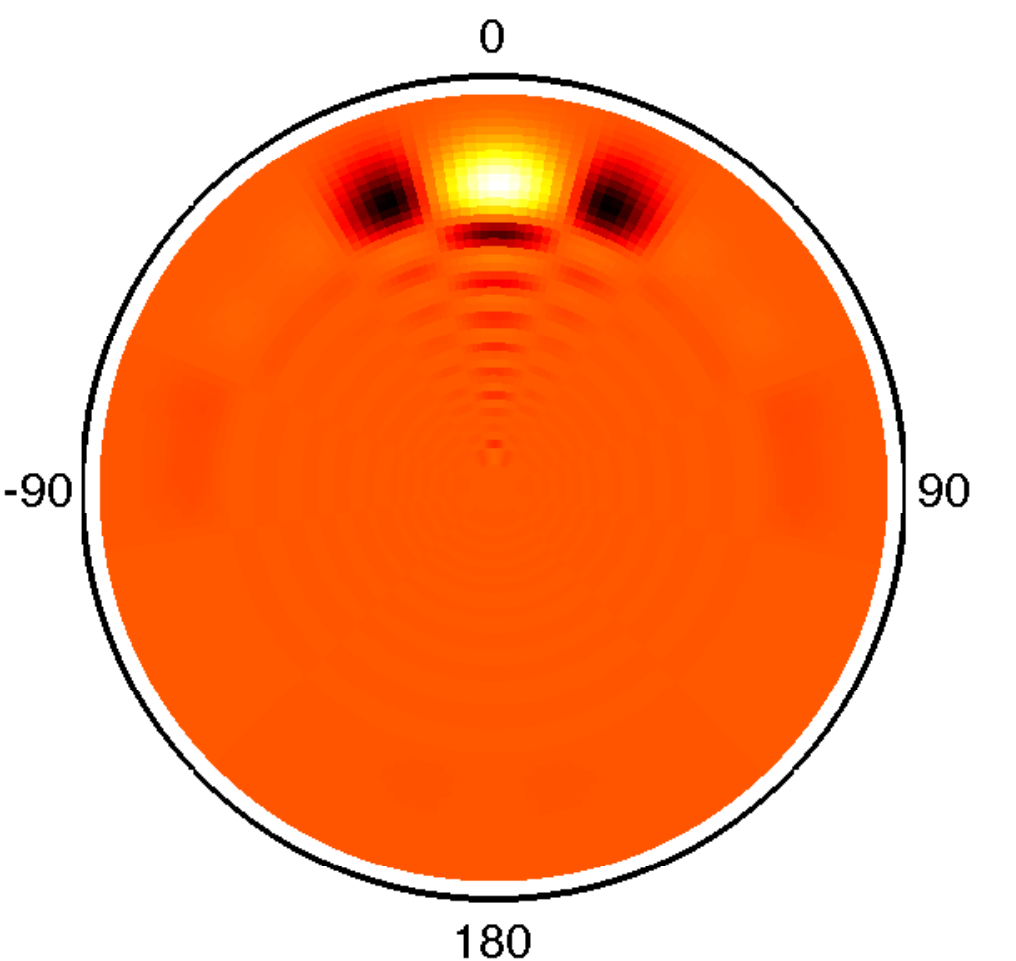}}
\subfigure[$(j, j^\prime) = (4,6)$]{\includegraphics[trim = 21.5cm 16.5cm 1.8cm 2cm, clip, width=4cm]{pics/wavelets.pdf}}	\\
\subfigure[$(j, j^\prime) = (5,5)$]{\includegraphics[trim = 10.cm 3cm 13.3cm 15.5cm, clip, width=4cm]{pics/wavelets.pdf}}
\subfigure[$(j, j^\prime) = (5,6)$]{\includegraphics[trim = 21.5cm 3cm 1.8cm 15.5cm, clip, width=4cm]{pics/wavelets.pdf}}
\caption{Slices of four successive axisymmetric flaglet wavelet kernels, probing different scales in angular and radial space. The flaglet parameters are $\lambda=\nu=2$ and the kernels are constructed at resolution $L=N=92$ on a ball of radius $R=1$. For visualisation purposes we show the flaglets corresponding to $j\in\{4,5\}$ and $j^\prime\in\{5,6\}$, translated to $r=0.3$ and zoomed on the range $r\in[0,0.4]$. Kernels of angular order $j=4$ (first row) probe large angular scales compared to those of order $j=5$ (second row). Similarly, kernels of radial order $j^\prime=5$ (first column) probe large radial scales compared to those of order $j^\prime=5$ (second column).}
\label{fig:wavelets}
\end{figure}

The flaglets and scaling function are constructed from their generating functions to satisfy the admissibility condition given by Eqn.~(\ref{identity}). A natural approach is to define ${\Psi}^{jj^\prime}_{\ell m p}$ from the generating functions $\kappa_\lambda$ and $\kappa_\nu$ to have support on $[\lambda^{j-1},\lambda^{j+1}]\times[\nu^{j^\prime-1},\nu^{j^\prime+1}]$, yielding
\begin{equation}
	{\Psi}^{jj^\prime}_{\ell m p} \equiv \sqrt{ \frac{2 \ell+1}{4\pi}}  \ \kappa_\lambda\left(\frac{\ell}{\lambda^j}\right) \kappa_\nu\left(\frac{\phantom{\ell}\hspace*{-2mm}p}{\nu^{j^\prime}}\right) \delta_{m0}.
\end{equation}
With these kernels, Eqn.~(\ref{identity}) is satisfied for $\ell > \lambda^{J_0}$ and \mbox{$p > \nu^{J^\prime_0}$}, where $J_0$ and $J^\prime_0$ are the lowest wavelet scales used in the decomposition. The scaling function $\Phi$ is constructed to extract the modes that cannot be probed by the flaglets:\footnote{Note that despite its piecewise definition ${\Phi}_{\ell m p}$ is continuous along and across the boundaries $p=\nu^{J_0^\prime}$ and $\ell = \lambda^{J_0}$. }
\begin{equation}
	{\Phi}_{\ell m p}  \equiv   \left\{ \begin{array}{ll} 
	 \hspace{-1mm} \sqrt{ \frac{2 \ell+1}{4\pi}}  \ \eta_{\nu}\left(\frac{\phantom{\ell}\hspace*{-2mm}p}{\nu^{J^\prime_0}}\right)\delta_{m0} , & \textrm{if } \ell  > \lambda^{J_0}, \ p \leq \nu^{J^\prime_0}	\nonumber \\
  \hspace{-1mm} \sqrt{ \frac{2 \ell+1}{4\pi}}  \ \eta_{\lambda}\left(\frac{\ell}{\lambda^{J_0}}\right)\delta_{m0}  , & \textrm{if } \ell \leq \lambda^{J_0} , \ p  > \nu^{J_0^\prime} \nonumber \\
\hspace{-1mm}  \sqrt{ \frac{2 \ell+1}{4\pi}}  \ \eta_{\lambda \nu}\left(\frac{\ell}{\lambda^{J_0}},\frac{p}{\nu^{J^\prime_0}}\right)\delta_{m0} ,  \hspace{-2mm} & \textrm{if }  \ell < \lambda^{J_0}, \ p  < \nu^{J_0^\prime} \\
  \quad 0 , & \textrm{elsewhere.} \end{array} \right.
\end{equation}
To satisfy exact reconstruction, $J$ and $J'$ are defined from the band-limits by $J = \lceil \log_\lambda(L-1) \rceil$ and \mbox{$J^\prime = \lceil \log_\nu(P-1) \rceil$}. The choice of $J_0$ and $J^\prime_0$ is arbitrary, provided that $0 \leq J_0 < J$ and $0 \leq J^\prime_0 < J^\prime$.  This framework generalises the notion of the harmonic tiling used to construct exact wavelets on the sphere \cite{Marinucci2008needlets, mcewen2008dirwavelets}; in fact, the flaglets defined here reduce in angular part to the wavelets defined in \cite{mcewen2008dirwavelets} for the axisymmetric case.  The flaglets and scaling function tiling of the \bl{Fourier-Laguerre} space of the ball is illustrated in Figure~\ref{fig:tiling}.  Flaglets and the scaling function may be reconstructed in the spatial domain from their harmonic coefficients. In Figure~\ref{fig:wavelets} flaglets are plotted in the spatial domain for a range of different scales; translated flaglets are plotted in Figure~\ref{fig:wavtransl}. The flaglets are \bl{well localised} in both real and \bl{Fourier-Laguerre} spaces and their angular aperture is invariant under radial translation.
 
\section{Multiresolution Algorithm} \label{sec:multiresolution}

In this section we discuss our implementation of the Fourier-Laguerre and flaglet transforms. We notably introduce a multiresolution algorithm for the flaglet transform to capture each wavelet scale in the minimal number of samples on the ball, thereby reducing the computational cost of the transform. We finally provide accuracy and complexity tests for our implementation of both transforms, which we make publicly available.

\subsection{Algorithm} \label{sec:algorithm}

In our framework, each flaglet ${\Psi}^{jj^\prime}$ has compact support in \bl{Fourier-Laguerre} space on $\ell \times p \in [\lambda^{j-1},\lambda^{j+1}]\times[\nu^{j^\prime-1},\nu^{j^\prime+1}]$, as shown in Figure~\ref{fig:tiling}. Thus, ${\Psi}^{jj^\prime}$ has band-limits in $\ell$ and $p$ of $\lambda^{j+1}$ and $\nu^{j^\prime+1}$ respectively.
For a band-limited function \mbox{$f\in L^2(B^3)$}, recall that the \mbox{$jj^\prime$-th} wavelet contribution is given by the simple product of Eqn.~(\ref{wav2}) in harmonic space.  Consequently, the band-limits of ${W}^{{\Psi}^{jj^\prime}}$ are given by the minimum of the band-limits of $f$ and ${\Psi}^{jj^\prime}$. Thus, for $j<J$ or $j^\prime<J^\prime$ the wavelet scale ${W}^{{\Psi}^{jj^\prime}}$ can be represented in fewer samples than $f$, without any loss of information. We exploit this property by designing a multiresolution approach where each wavelet scale is represented in real space with the smallest number of samples necessary. Note that the scaling function must be used at full resolution since its angular and radial band-limits are $L$ and $P$ respectively.  To summarise the multiresolution algorithm, although $f$ is decomposed at full resolution, the wavelets coefficients are reconstructed in real space with the minimum number of samples supporting their band-limits. This leads to a significant reduction in computation time, which is then dominated by the small number of full resolution Fourier-Laguerre transforms. 

\subsection{\bl{Fast} implementation} \label{sec:implementation}

Our implementation of the algorithms of this article \bl{is made available in the following three packages}, which are written in {\tt C} and include {\tt MATLAB} interfaces for most high-level features, and are described in turn:
\begin{itemize}
\item {\tt FLAG}: spherical Laguerre transform and Fourier-Laguerre transforms on the ball (exact spherical Bessel and Fourier-Bessel decompositions are optional features that additionally require the {\tt GNU} Math Library\footnote{http://www.gnu.org/software/gsl/}).
\item {\tt S2LET}: axisymmetric wavelet transform on the sphere through harmonic tiling.
\item {\tt FLAGLET}: axisymmetric flaglet transform on the ball, combining {\tt FLAG} and {\tt S2LET} to construct flaglets in Fourier-Laguerre space through harmonic tiling.
\end{itemize} 
We make these three packages publicly available.\footnote{\bl{\url{http://www.flaglets.org/}}}
All packages require {\tt SSHT}\footnote{\url{http://www.jasonmcewen.org/}}, which implements fast and exact algorithms to perform the forward and inverse spherical harmonic transforms corresponding to the MW sampling theorem \cite{mcewen2011novelsampling}. {\tt SSHT} requires the {\tt FFTW}\footnote{\url{http://www.fftw.org/}} package.

Since the naive spherical harmonic transform scales as $\mathcal{O}(L^4)$ and the spherical Laguerre transform scales as $\mathcal{O}(P^2)$, the naive complexity of the Fourier-Laguerre transform is $\mathcal{O}(P^2L^4)$.  However, rather than computing triple integrals/sums over the ball directly, it is straightforward to show that the Fourier-Laguerre transform can be performed separately on the sphere and on the radial half-line, like the Fourier-Bessel transform \cite{leistedt20113dex}. Since the angular and radial samplings are separable, the related transforms can be computed independently through a separation of variables, \bl{so that the complexity reduces to $\mathcal{O}(Q^5)$ for $Q \sim P \sim L$}.
The separation of variables also means we are able to exploit high-performance recurrences and algorithms that exist for both the spherical Laguerre and spherical harmonic transforms.  In particular, the radial basis functions $K_p(r)$ are calculated using a normalised recurrence formula derived from the recurrence on the Laguerre polynomials. Moreover, a critical point for the accuracy of the Fourier-Laguerre transform is the computation of the Gauss-Laguerre quadrature, for which we use the previous normalised recurrence complemented with an appropriate root-finder algorithm.  The fast spherical harmonic transforms implemented in the {\tt SSHT} package use the Trapani \& Navaza method \cite{trapani2006} to efficiently compute Wigner functions (which are closely related to the spherical harmonics) through recursion.\footnote{Alternatively, Risbo's method could also be used \bl{to} compute Wigner functions \cite{risbo1996}.}  These fast spherical harmonic transform algorithms \cite{mcewen2011novelsampling} scale as $\mathcal{O}(L^3)$. The final complexity achieved by the Fourier-Laguerre transform is thus $\mathcal{O}(\bl{Q^4})$. 

The flaglet transform (forward and inverse) is calculated in a straightforward manner in \bl{Fourier-Laguerre} space, thus its computation is dominated by the Fourier-Laguerre transform of the signal, approximation coefficients, and wavelets coefficients at all scales, requiring \mbox{$[(J+1-J_0)(J^\prime+1-J_0^\prime)+2]$} Fourier-Laguerre transforms. If all wavelet contributions are reconstructed at full resolution in real space, the overall wavelet transform scales as $\mathcal{O}( [(J+1-J_0)(J^\prime+1-J_0^\prime){+2}]\bl{Q^4})$. Note that $J$ and $J^\prime$ depend on the band-limits $L$ and $P$ and the parameters $\lambda$ and $\nu$, respectively.  However, in the previous section we established a multiresolution algorithm that takes advantage of the band-limits of the individual flaglets.  With this algorithm, only the scaling function and the finest wavelet scales (\emph{i.e.} $j\in\{J-1,J\}$ and $j^\prime\in\{J^\prime-1,J^\prime\}$) are computed at maximal resolution corresponding to band-limits $L$ and $P$. The complexity of the overall multiresolution flaglet transform is then dominated by these operations and scales as $\mathcal{O}(\bl{Q^4})$.

\subsection{Numerical validation} \label{sec:numerical}

In this section we evaluate {\tt FLAG} and {\tt FLAGLET} in terms of accuracy and complexity. We show that they achieve \bl{floating-point} precision and scale as detailed in the previous section. In both cases we consider band-limits \mbox{$L=P=2^i$} with \mbox{$i \in \{2,\ldots,9\}$} and generate sets of harmonic coefficients $f_{\ell m p}$ following independent Gaussian distributions $\mathcal{N}(0,1)$. We then perform either the Fourier-Laguerre or the flaglet decomposition, before reconstructing the harmonic coefficients, therefore denoted by $f_{\ell m p}^\textrm{rec}$.  We evaluate the accuracy of the transforms using the error metric \mbox{$\epsilon = \max | f_{\ell m p} - f_{\ell m p}^\textrm{rec}|$}, which is theoretically zero for both transforms since all signals are band-limited by construction. The complexity is quantified by observing how the computation time \mbox{$t_{\rm c} = [t_{\textrm{synthesis}} +  t_{\textrm{analysis}} ]/2$} scales with the band-limits, where the synthesis and analysis computation times, $t_{\textrm{synthesis}}$ and $t_{\textrm{analysis}}$ respectively, are defined explicitly for the two transforms in the paragraphs that follow. The stability of both $\epsilon$ and $t_{\rm c}$ is checked by averaging over hundreds of realisations of $f_{\ell m p}$ in the cases $i\in \{2,\dots,7\}$ and a small number of realisations for $i\in\{8,9\}$. Recall that for given band-limits $L$ and $P$ the number of samples on the ball required by the exact quadrature is $N = P[(2L-1)(L-1)+1]$. {All tests were run on a 2.5GHz Core i5 processor with 8GB of RAM.} 

\begin{figure}[]\centering
\subfigure[Numerical accuracy of the Fourier-Laguerre transform]{\includegraphics[trim = 7.4cm 15cm 1cm 1cm, clip, width=9cm]{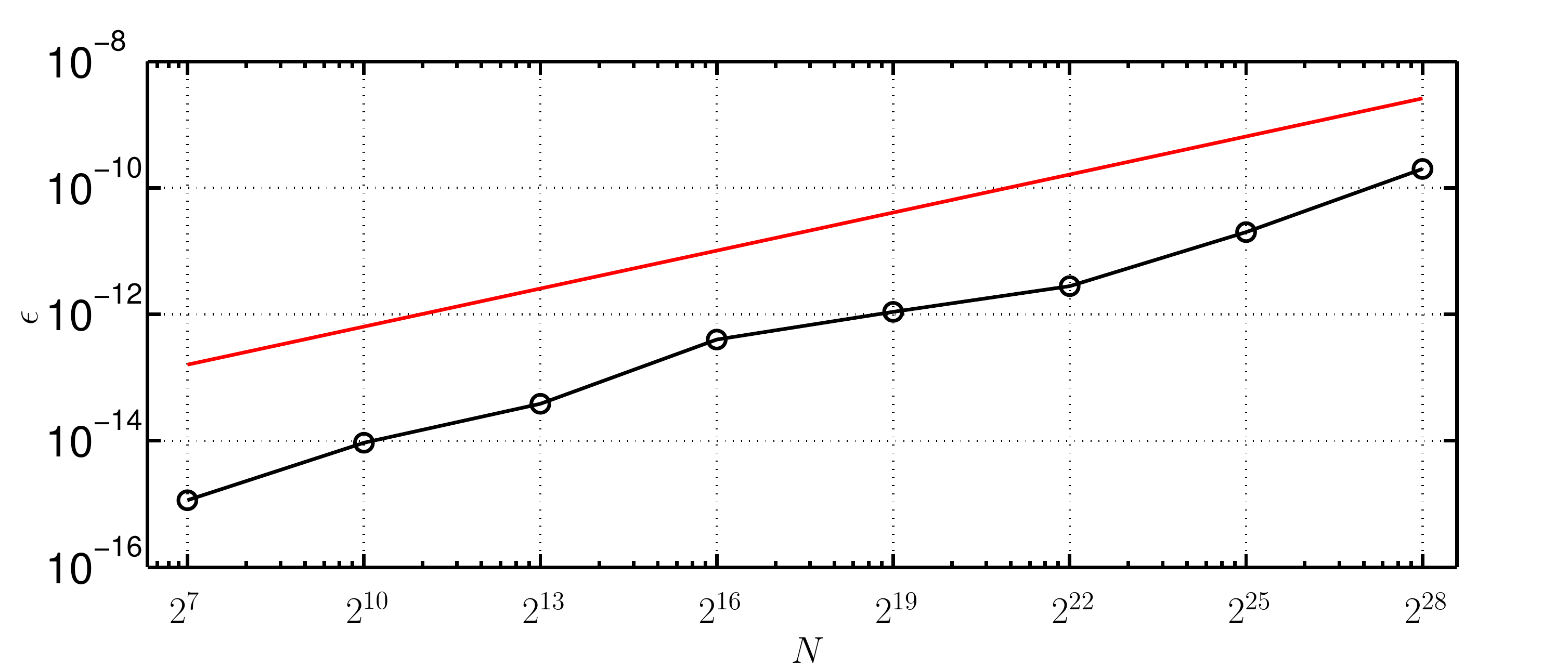}}
\subfigure[Computation time of the Fourier-Laguerre transform]{\includegraphics[trim = 7.4cm 1.cm 1cm 15cm, clip, width=9cm]{pics/flagperfs.pdf}}
\caption{Numerical accuracy and computation time of the Fourier-Laguerre transform computed with {\tt FLAG}, where $N$ corresponds to the number of samples on the ball required to capture all the information contained in the band-limited test signal. We consider $L=P=2^i$ with $i \in \{2,\ldots,9\}$. These results are averaged over many realisations of random band-limited signals and were found to be very stable.  Very good numerical accuracy is achieved, with numerical errors comparable to \bl{floating-point} precision, found empirically to scale as $\mathcal{O}(\bl{Q^2})$ as shown by the red line in panel~(a), \bl{where $Q \sim P \sim L$}. Computation time scales as $\mathcal{O}(\bl{Q^4})$ as shown by the red line in panel~(b), in agreement with theory.}
\label{fig:flagperfs}
\end{figure}

The results of these tests for the Fourier-Laguerre transform are presented on Figure~\ref{fig:flagperfs}. The indicators $\epsilon$ and $t_{\rm c}$ are plotted against the number of samples $N$. Each test starts from coefficients $f_{\ell m p}$ randomly generated. The synthesis refers to constructing the band-limited signal $f$ from the decomposition $f_{\ell m p}$. The analysis then corresponds to decomposing $f$ into Fourier-Laguerre coefficients $f^{\textrm{rec}}_{\ell m p}$. As shown in Figure~\ref{fig:flagperfs}, {\tt FLAG} achieves very good numerical accuracy, with numerical errors comparable to \bl{floating-point} precision, and computation time scales as $\mathcal{O}(\bl{Q^4})$, in agreement with theory.  

\begin{figure}[]\centering
\subfigure[Numerical accuracy of the flaglet transform]{\includegraphics[trim = 7.4cm 14.5cm 1cm 1cm, clip, width=9cm]{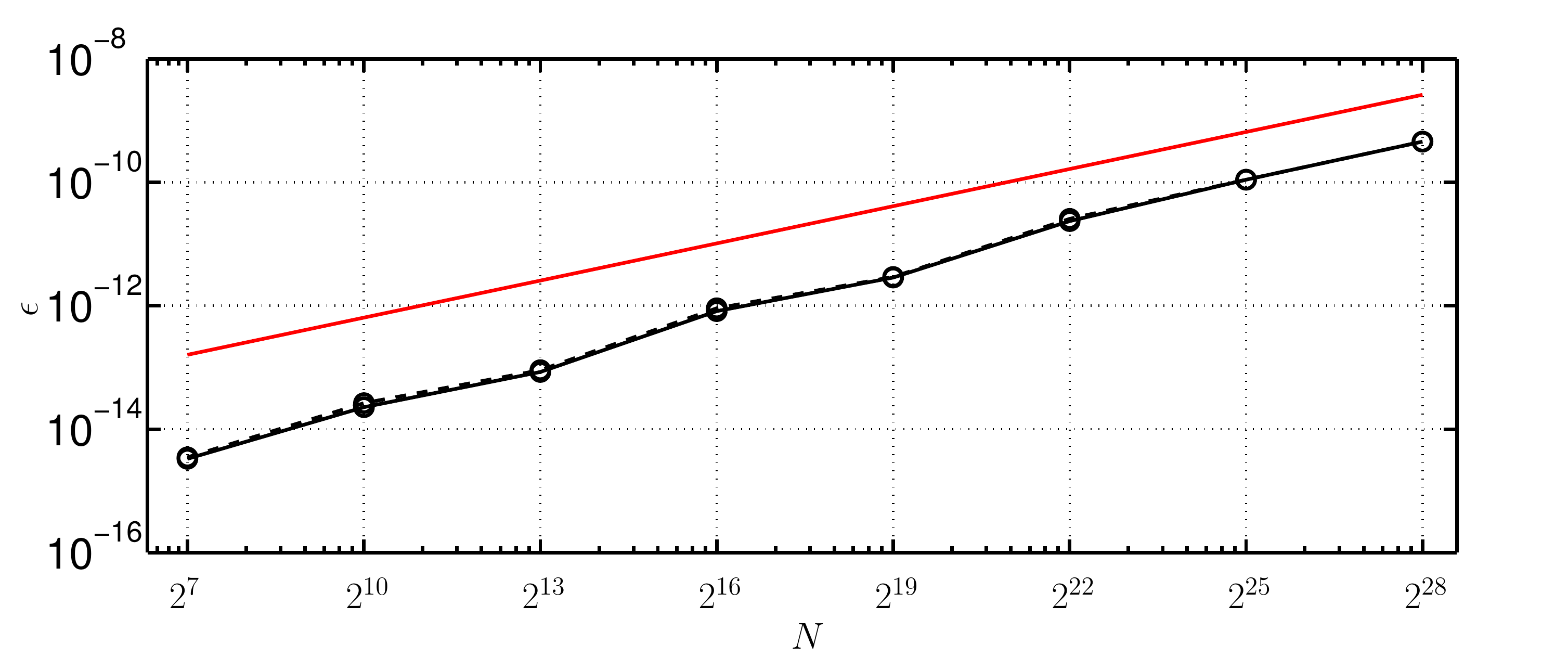}}
\subfigure[Computation time of the flaglet transform]{\includegraphics[trim = 7.4cm 1.cm 1cm 15cm, clip, width=9cm]{pics/b3letperfs.pdf}}
\caption{Numerical accuracy and computation time of the flaglet transform computed with {\tt FLAGLET}, where $N$ corresponds to the number of samples on the ball required to capture all the information contained in the band-limited test signal.  We consider $L=P=2^i$ with $i \in \{2,\ldots,9\}$, with parameters $\lambda = \nu = 2$, $J_0 = J_0^\prime=0$. These results are averaged over many realisations  of random band-limited signals and were found to be very stable. The flaglet transform is either performed at full-resolution (dashed lines) or with the multiresolution algorithm (solid lines). Very good numerical accuracy is achieved by both the full resolution and multiresoltion algorithms (which achieve indistinguishable accuracy), with numerical errors comparable to \bl{floating-point} precision, found empirically to scale as $\mathcal{O}(PL)$ as shown by the red line in panel~(a).  The multiresolution algorithm is ten times faster than the full-resolution approach.  Computation time scales as $\mathcal{O}(PL^3)$ for both algorithms as shown by the red line in panel~(b), in agreement with theory.}
\label{fig:b3letperfs}
\end{figure}

The results of similar tests for the flaglet transform (entirely performed in real space) are presented on Figure~\ref{fig:b3letperfs}. As previously, the indicators $\epsilon$ and $t_{\rm c}$ are plotted against the number of samples $N$. Since we evaluate the flaglet transform in real space, a preliminary step is required to construct a band-limited signal $f$ from the randomly generated $f_{\ell m p}$. This step is not included in the computation time since its only purpose is to generate a valid band-limited test signal in real space. The analysis then denotes the decomposition of $f$ into wavelet coefficients $W^{{\Psi}^{jj^\prime}}$ and scaling coefficients $W^\Phi$ on the ball. The synthesis refers to recovering the signal $f^{\textrm{rec}}$ from these coefficients. The final step, which is not included in the computation time, is to decompose $f^{\textrm{rec}}$ into Fourier-Laguerre coefficients $f^{\textrm{rec}}_{\ell m p}$ in order to compare them with $f_{\ell m p}$. As shown in Figure~\ref{fig:b3letperfs}, {\tt FLAGLET} achieves very good numerical accuracy, with numerical errors comparable to \bl{floating-point} precision. Moreover, the full resolution and multiresolution algorithms are indistinguishable in terms of accuracy. However, the latter is ten times faster than the former since only the scaling function and a small number of wavelet coefficients are computed at full resolution.  As shown in Figure~\ref{fig:b3letperfs}, computation time scales as $\mathcal{O}(\bl{Q^4})$ for both algorithms, in agreement with theory.
 
\section{Denoising Illustration}\label{sec:denoisingexample}

In this section we illustrate the use of the flaglet transform in the context of a simple denoising problem. We consider two datasets naturally defined on the ball and contaminate them with band-limited noise. We compute the flaglet transform of the noisy signal and perform simple denoising by hard-threshold the wavelet coefficients.  We reconstruct the signal from the thresholded wavelet coefficients and examine the improvement in signal fidelity.

\subsection{Wavelet denoising} 

Consider the noisy signal $y = s+n \in L^2(B^3)$, where the signal of interest $s\in L^2(B^3)$ is contaminated with noise $n\in L^2(B^3)$. A simple way to evaluate the fidelity of the observed signal $y$ is to examine the signal-to-noise ratio, which we define on the ball by
\begin{equation}
	\textrm{SNR}(y) \equiv 10 \log_{10} \frac{ \| s \|_2^2 }{ \| y - s \|_2^2 }.
\end{equation}
The signal energy is given by 
\begin{equation}
	\| y \|_2^2 \ \equiv \  \langle y|y \rangle= \ \int_{B^3} {\rm d}^3\vect{r}  | y(\vect{r}) |^2  \ = \ \sum_{\ell m p } | y_{\ell m p} |^2,
\end{equation}
where the final equality follows from a Parseval relation on the ball (which follows directly from the orthogonality of the Fourier-Laguerre basis functions).  In practice, we compute signal energies through the final \bl{Fourier-Laguerre} space expression to avoid the necessity of an explicit quadrature rule.

We seek a denoised version of $y$, denoted by $d \in L^2(B^3)$, such that ${\rm SNR}(d)$ is as large as possible in order to extract the informative signal $s$.  We take the flaglet transform of the noisy signal since we intend to denoise the signal in wavelet space, where we expect the energy of the informative signal to be concentrated in a small number of wavelet coefficients while the noise energy will be spread over many wavelet coefficients.  Since the flaglet transform is linear, the wavelet coefficients of the $jj^\prime$-th scale of the noisy signal is simply the sum of the individual contributions:
\begin{equation}
	Y^{j j^\prime}(\vect{r}) = S^{j j^\prime}(\vect{r}) +  N^{j j^\prime} (\vect{r}),
\end{equation}
where capital letters denote the wavelet coefficients, \emph{i.e.} \mbox{$Y^{{jj^\prime}} \equiv y \star \Psi^{jj^\prime}$}, $S^{{jj^\prime}} \equiv s \star \Psi^{jj^\prime}$ and $N^{{jj^\prime}} \equiv n \star \Psi^{jj^\prime}$.

In the illustrations performed here, we assume the noise model
\begin{equation}
	 \mathbb{E}\left( |{n}_{\ell m p}|^2 \right) \ = \ \sigma^2  \left(\frac{p}{P}\right)^2 \delta_{\ell \ell ^\prime} \delta_{mm^\prime} \delta_{pp^\prime}, \label{noisevar}
\end{equation}
which corresponds to a \bl{white noise} for the angular space with a dependence on the radial mode $p$, where $\mathbb{E}(\cdot)$ denotes ensemble averages. We do not opt for a \bl{white noise} in radial space (\emph{i.e.} $ \mathbb{E}\left( |{n}_{\ell m p}|^2  \right) = \sigma^2 \delta_{\ell \ell ^\prime} \delta_{mm^\prime} \delta_{pp^\prime}$) because the latter has its energy concentrated in the centre of the ball due to the shape of the spherical Laguerre basis functions. The $p$-dependence gives a greater weight to small-scale radial features and hence yields a more homogeneous noise on the ball, which is more useful for visualisation purposes. For this noise model one can show that the expected covariance of the wavelet coefficients of the $jj^\prime$-th scale reads
\begin{eqnarray}
	\mathbb{E}\left( |{N}^{j j^\prime}\hspace{-1mm}(r,\omega)|^2 \right) &=& \sigma^2 \sum_{\ell p} \left(\frac{p}{P}\right)^2 |  {\Psi}^{jj^\prime}_{\ell 0 p} |^2 |K_p(r)|^2   \ \ \label{noisemodel} \\ 
	 &\equiv & \left( \sigma^{jj^\prime}\hspace{-1mm}(r) \right)^2. \nonumber
\end{eqnarray}

Denoising is performed by hard-thresholding the wavelet coefficients $Y^{j j^\prime}$, where the threshold is taken as \mbox{$T(\vect{r}) = 3\sigma^{jj^\prime}(r)$}.  The wavelet coefficients of the denoised signal \mbox{$D^{{jj^\prime}} \equiv d \star \Psi^{jj^\prime}$} are thus given by
\begin{equation}
	D^{j j^\prime}(\vect{r}) = \left\{\begin{array}{ll}
		0 ,   & \textrm{if } Y^{j j^\prime}(\vect{r})< T(\vect{r}) \\
		Y^{j j^\prime}(\vect{r}),    &\textrm{otherwise.}
		\end{array}\right. \label{threshold}
\end{equation}
The denoised signal $d$ is then reconstructed from its wavelet coefficients and scaling coefficients (the latter are not thresholded and thus not altered). To assess the effectiveness of this simple flaglet denoising strategy when the informative signal $s$ is known, we compute the SNR of the denoised signal and compare it to the SNR of the original noisy signal.  In what follows we apply this simple denoising technique to two datasets naturally defined on the ball.

\subsection{Examples} 

\begin{figure}[]\centering
\subfigure[Band-limited data]{\includegraphics[trim = 2cm 22cm 10.6cm 1cm, clip, width=4.3cm]{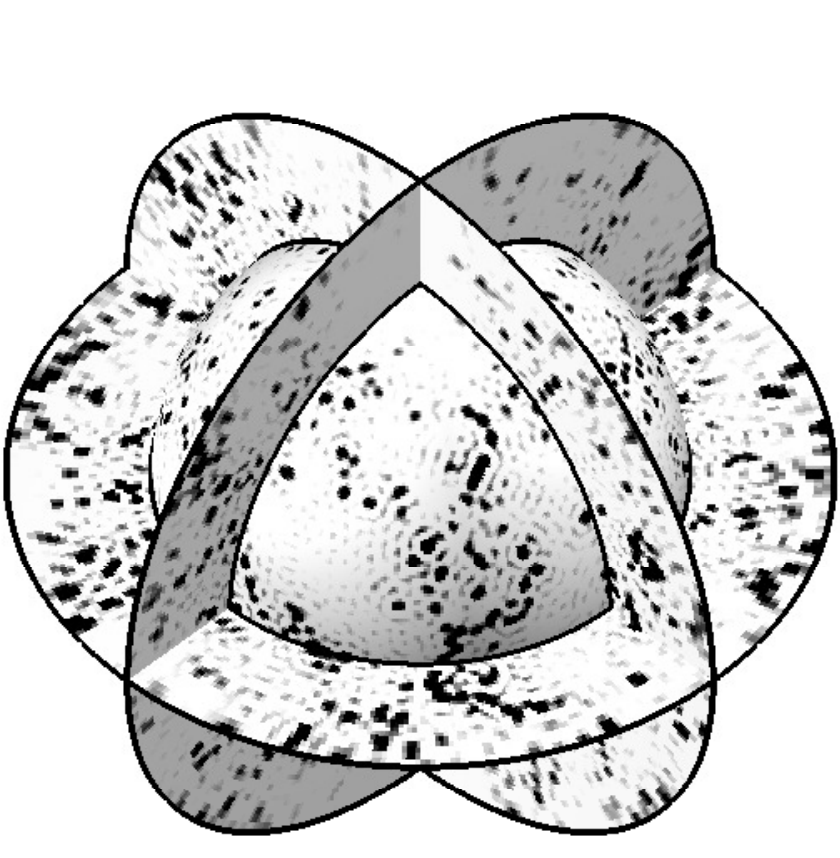}}
\subfigure[Scaling coefficients]{\includegraphics[trim = 11.3cm 22cm 1.3cm 1cm, clip, width=4.3cm]{pics/lsswav.pdf}}
\\
\subfigure[$(j,j^\prime)=(6,6)$]{\includegraphics[trim = 2cm 12.5cm 10.6cm 11.2cm, clip, width=4.3cm]{pics/lsswav.pdf}}
\subfigure[$(j,j^\prime)=(6,7)$]{\includegraphics[trim = 11.3cm 12.5cm 1.3cm 11.2cm, clip, width=4.3cm]{pics/lsswav.pdf}}
\\
\subfigure[$(j,j^\prime)=(7,6)$]{\includegraphics[trim = 2cm 3.0cm 10.6cm 21.1cm, clip, width=4.3cm]{pics/lsswav.pdf}}
\subfigure[$(j,j^\prime)=(7,7)$]{\includegraphics[trim = 11.3cm 3.0cm 1.3cm 21.1cm, clip, width=4.3cm]{pics/lsswav.pdf}}
\caption{Flaglet decomposition of the N-body simulation dataset considered for the first denoising example. The initial dataset was pixelised and band-limited at $L=P=128$. The flaglet parameters are $\lambda=\nu=2$ (giving $J=J^\prime=7$) and the scaling coefficients correspond to $J_0=J_0^\prime=6$ since the lower scale indices do not contain a great deal of information. The four wavelet coefficients together with the scaling coefficients decompose the initial dataset exactly, \emph{i.e.} the original signal can be recovered perfectly from these wavelet and scaling coefficients. All signals were oversampled on $L=P=256$ for visualisation purposes.}
\label{fig:lsswav}
\end{figure}

\begin{figure}[]\centering
\subfigure[Band-limited data]{\includegraphics[trim = 3.9cm 16.3cm 18.1cm 0cm, clip, width=4.3cm]{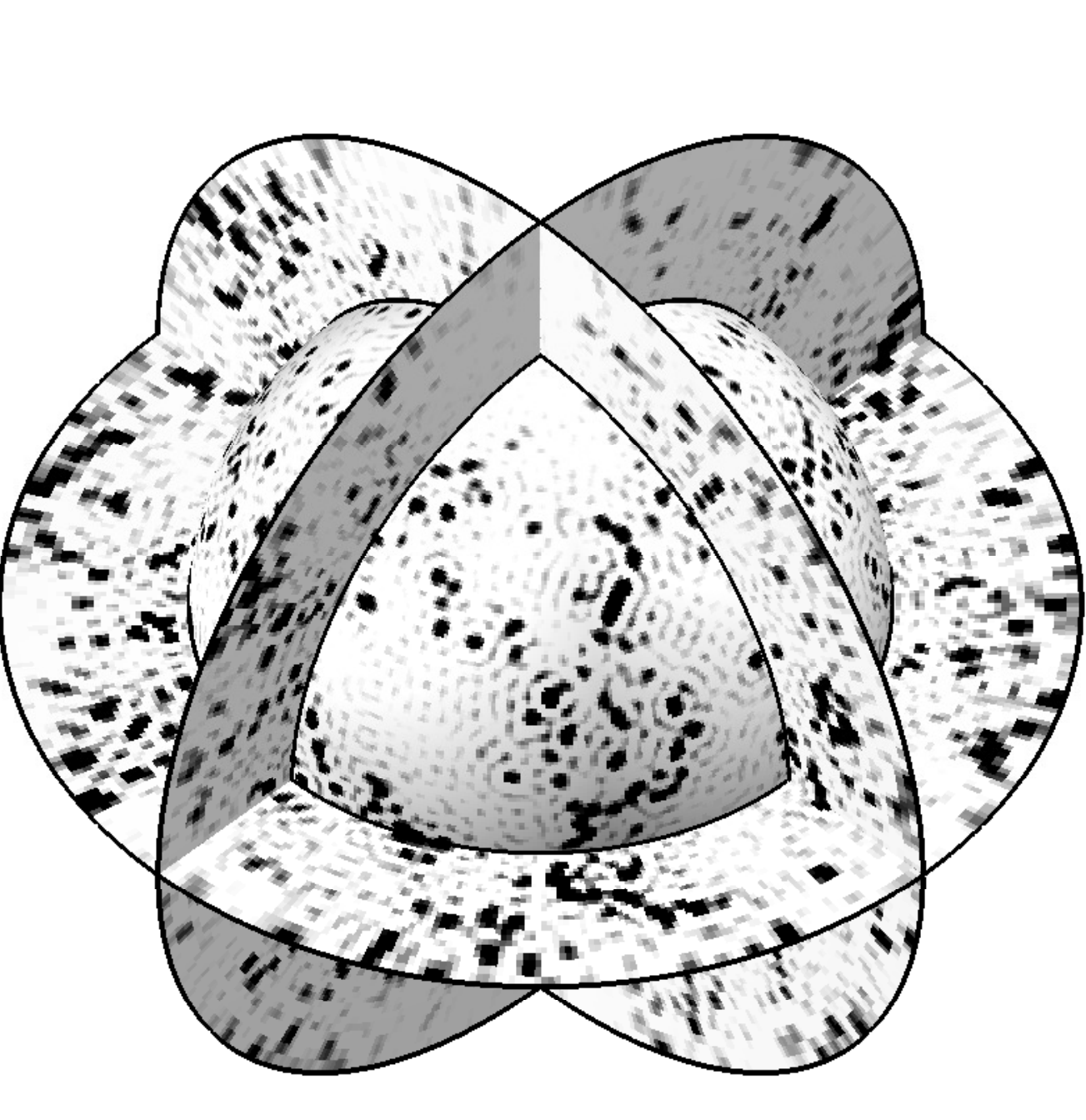}}
\subfigure[Noise]{\includegraphics[trim = 19.4cm 16.3cm 2.6cm 0cm, clip, width=4.3cm]{pics/lss_5_12.pdf}}
\\
\subfigure[Noisy signal]{\includegraphics[trim = 3.9cm 2.3cm 18.1cm 15cm, clip, width=4.3cm]{pics/lss_5_12.pdf}}
\subfigure[Denoised signal]{\includegraphics[trim = 19.4cm 2.3cm 2.6cm 14.5cm, clip, width=4.3cm]{pics/lss_5_12.pdf}}
\caption{Denoising of an N-body simulation. The data are contaminated with a band-limited noise and decomposed into wavelet coefficients. Denoising is performed by a simple hard-thresholding of the wavelet coefficients, following a noise model. The denoised signal is then reconstructed from the thresholded wavelet coefficients. In this example, for an initial SNR of $5$dB, the flaglet denoised signal is recovered with $\textrm{SNR}$ of $\textrm{SNR}=11$dB (with resolution $L=P=128$, oversampled on $L=P=256$ and using flaglet parameters $\lambda = \nu = 2$, $J_0=J^\prime_0=0$, giving $J=J^\prime=7$).}
\label{fig:denoising_lss}
\end{figure}

The first dataset we consider is the full-sky Horizon simulation \cite{teyssier2009horizon}: an N-body simulation covering a 1Gpc periodic box of 70 billion dark matter particles generated from the concordance model cosmology derived from 3-year Wilkinson Microwave Anisotropy Probe (WMAP) observations \cite{wmap3Spergel}. The purpose of such a simulation is to reproduce the action of gravity (and to a minor extent galaxy formation) on a large system of particles, with the initial conditions drawn from a cosmological model of interest. The outcome is commonly used to confront astrophysical models with observations. For simplicity we only consider a ball of 1MPc radius centered at the origin so that the structures are of reasonable size. Figure~\ref{fig:lsswav} shows the initial data, band-limited at $L=P=128$, as well as their wavelet coefficients with $\lambda = \nu = 2$, $J=J^\prime=7$ and scaling coefficients for $J_0=J_0^\prime=6$ since the lower scale indices do not contain a great deal of information. We see that the filamentary distribution of matter is naturally suited to a flaglet analysis on the ball since the informative signal is likely to be contained in a reduced number of wavelet coefficients.
The original data are corrupted by the addition of random noise defined by Eqn.~(\ref{noisevar}) for an SNR of $5$dB.  The wavelet denoising procedure described previously is then applied.  The denoised signal is recovered with an SNR of $11$dB, highlighting the effectiveness of this very simple flaglet denoising strategy on the ball. The results of this denoising illustration are presented in Figure~\ref{fig:denoising_lss}.

\begin{figure}[]\centering
\subfigure[Band-limited data]{\includegraphics[trim = 3.9cm 16.3cm 18.1cm 0cm, clip, width=4.3cm]{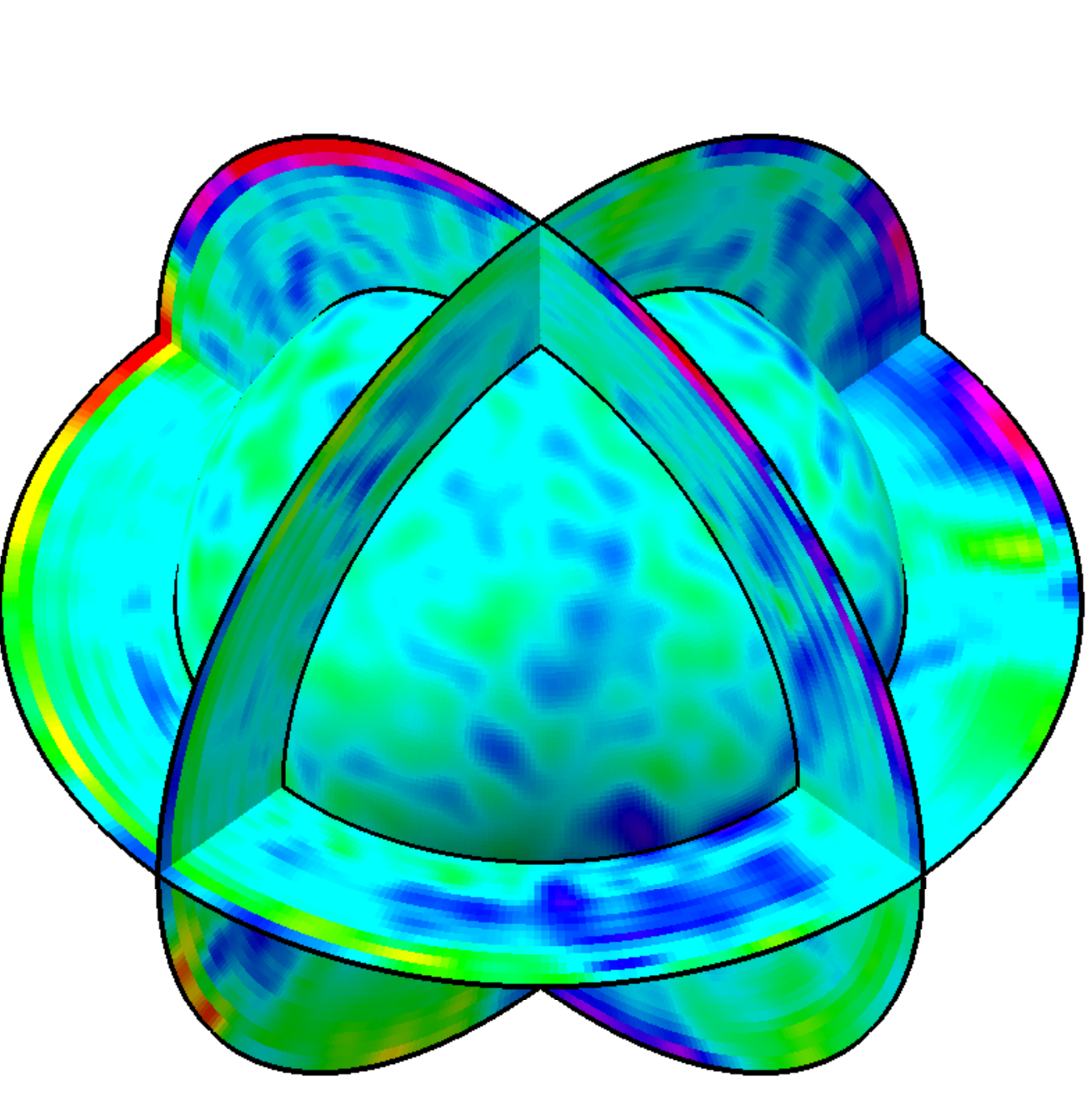}}
\subfigure[Noise]{\includegraphics[trim =  19.4cm 16.3cm 2.6cm 0cm, clip, width=4.3cm]{pics/geo_5_17.pdf}}
\\
\subfigure[Noisy signal]{\includegraphics[trim =  3.9cm 2.3cm 18.1cm 15cm, clip, width=4.3cm]{pics/geo_5_17.pdf}}
\subfigure[Denoised signal]{\includegraphics[trim = 19.4cm 2.3cm 2.6cm 14.5cm, clip, width=4.3cm]{pics/geo_5_17.pdf}}
\caption{Denoising of a seismological Earth model. The data are contaminated with a band-limited noise and decomposed into wavelet coefficients. Denoising is performed by a simple hard-thresholding of the wavelet coefficients, following a noise model. The denoised signal is then reconstructed from the thresholded wavelet coefficients.  In this example, for an initial SNR of $5$dB, the flaglet denoised signal is recovered with $\textrm{SNR}$ of $17$dB (with resolution $L=P=128$ and using flaglet parameters $\lambda = \nu = 3$, $J_0=J^\prime_0=0$, giving $J=J^\prime=7$).}
\label{fig:denoising_geo}
\end{figure}

\bl{The second dataset we consider is Ritsema's seismological Earth model of shear wavespeed perturbations in the mantle,  known as S40RTS} \cite{ristema2010, simons2011, simons2011wavelets}.\footnote{\url{http://www.earth.lsa.umich.edu/~jritsema/}}  The model supplies spherical harmonic coefficients in the angular dimension and radial spline coefficients in the depth dimension to define a signal on the ball, which we band-limit.  Contrarily to the first example, Ritsema's model does not contain a lot of structure at the smallest scales but essentially contains large-scale features. 
As previously, the original data are corrupted by the addition of random noise defined by Eqn.~(\ref{noisevar}) for an SNR of $5$dB.  The flaglet denoising procedure described previously is then applied.  The denoised signal is recovered with an SNR of $17$dB, again highlighting the effectiveness of this very simple flaglet denoising strategy on the ball.  As expected, the improvement in SNR is better than for the previous dataset since the informative signal is mainly captured by a few large wavelet scales.  The results of this denoising illustration are presented in Figure~\ref{fig:denoising_geo}.

\section{Conclusions}\label{sec:conclusion}

One reason an exact wavelet transform of a band-limited signal on the ball has not yet been derived is due to the absence of an exact harmonic transform on the ball.  We have taken advantage of the orthogonality of the Laguerre polynomials on $\mathbb{R}^+$ to define the spherical Laguerre transform, a novel radial transform that admits an exact quadrature rule.  Combined with the spherical harmonics, we used this to derive a sampling theorem and an exact harmonic transform on the ball, which we call the Fourier-Laguerre transform. A function that is band-limited in Fourier-Laguerre space can be decomposed and reconstructed at \bl{floating-point} precision, and its Fourier-Bessel transform can be calculated exactly. For radial and angular band-limits $P$ and $L$, respectively, the sampling theorem guarantees that all the information of the band-limited signal is captured in a finite set of $N=P[(2L-1)(L-1)+1]$ samples on the ball. 

We have developed an exact wavelet transform on the ball, the so-called flaglet transform, through a tiling of the Fourier-Laguerre space. The resulting flaglets \bl{form a tight frame and are well localised in both real and \bl{Fourier-Laguerre} spaces. Their angular aperture is invariant under radial translation.}  We furthermore established a multiresolution algorithm to compute the flaglet transform, capturing all the information contained in each wavelet scale in the minimal number of samples on the ball, thereby reducing the computation cost of the flaglet transform considerably. 

Flaglets are a promising new tool for analysing signals on the ball, particularly for extracting spatially localised features at different scales of interest. Exactness of both the Fourier-Laguerre and the flaglet transforms guarantees that any band-limited signal can be analysed and decomposed into wavelet coefficients and then reconstructed without any loss of information. To illustrate these capabilities, we considered the denoising of two different datasets which were contaminated with synthetic noise.  A very simple flaglet denoising strategy was performed by hard-thresholding the wavelet coefficients of the noisy signal, before reconstructing the denoised signal from the thresholded wavelet coefficients.  In these illustrations a considerable improvement in SNR was realised by this simple flaglet denoising strategy, demonstrating the effectiveness of flaglets for the analysis of data defined on the ball.  Our implementation of all of the transforms and examples detailed in this article is made publicly available. In future work we intend to revoke the axisymmetric constraint by developing directional flaglets.

\bibliographystyle{IEEEtran}
\bibliography{biblio}
 

\begin{biography}[{\includegraphics[width=1in,height=1.25in,clip,keepaspectratio]{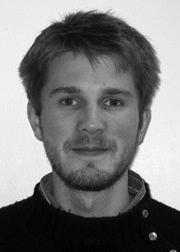}}]{Boris Leistedt}
received a Master's degree in Electrical Engineering jointly from the University of Mons, Belgium, and Sup\'elec, France, in 2011 as well as a M.Sc. in Physics from University Orsay Paris-Sud. He is currently a Ph.D. candidate in the Cosmology Group at University College London. 

His research interests span observational cosmology, inflationary physics and innovative methods to look for the imprints of the early universe on cosmological observables. 
\end{biography}

 \begin{biography}[{\includegraphics[width=1in,height=1.25in,clip,keepaspectratio]
 {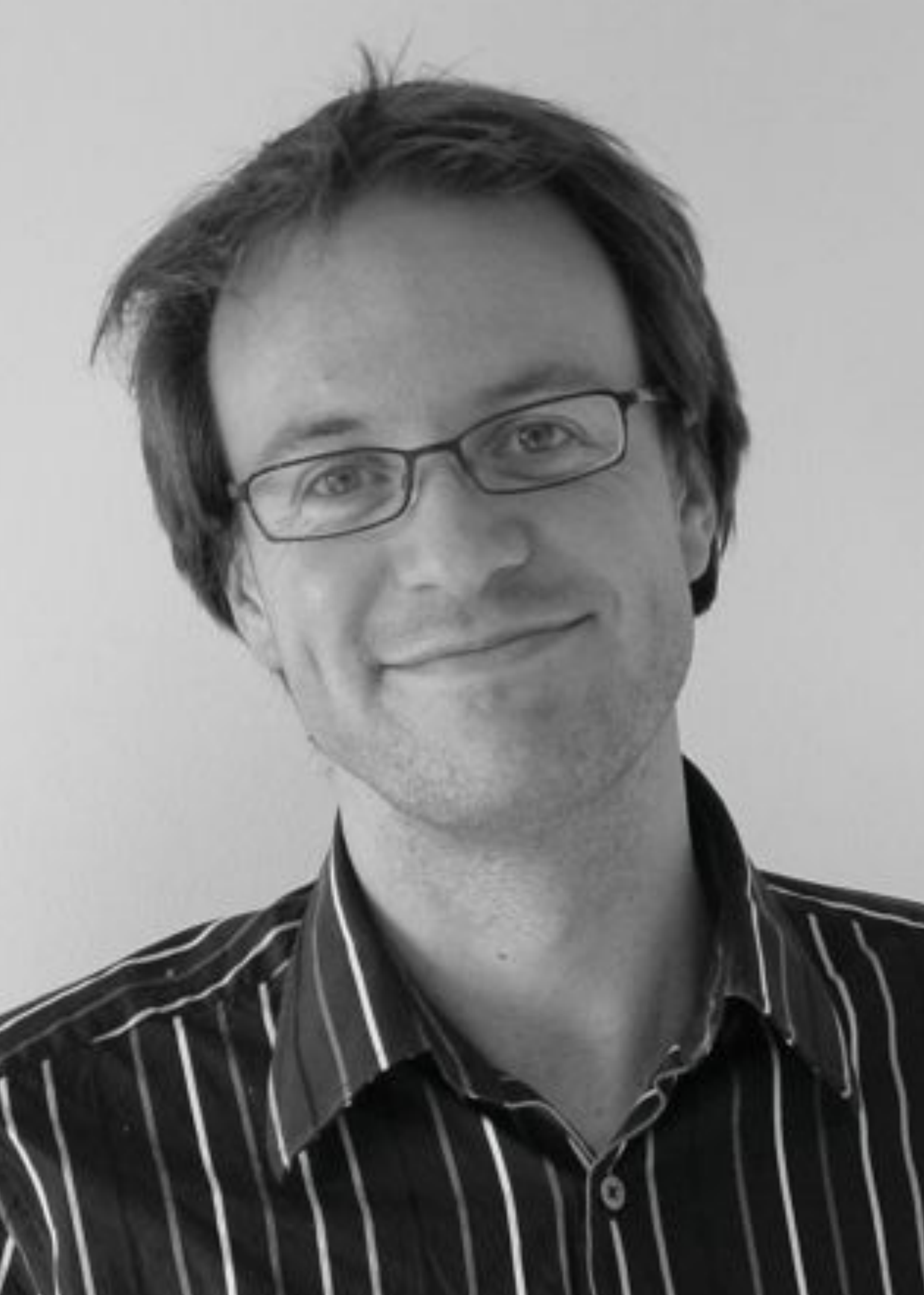}}]{Jason McEwen}
received a B.E.\ (Hons) degree in Electrical and Computer
  Engineering from the University of Canterbury, New Zealand, in 2002
  and a Ph.D.\ degree in Astrophysics from the University of Cambridge
  in 2007.

  He held a Research Fellowship at Clare College, Cambridge, from 2007
  to 2008, worked as a Quantitative Analyst from 2008 to 2010, and
  held a position as a Postdoctoral Researcher at Ecole Polytechnique
  F{\'e}d{\'e}rale de Lausanne (EPFL), Switzerland, from 2010 to
  2011. From 2011 to 2012 he held a Leverhulme Trust Early Career
  Fellowship at University College London (UCL), where he remains as a
  Newton International Fellow, supported by the Royal Society and the
  British Academy.  His research interests are focused on spherical
  signal processing, including sampling theorems and wavelets on the
  sphere, compressed sensing and Bayesian statistics, and applications
  of these theories to cosmology and radio interferometry.
\end{biography}

\end{document}